\pdfoutput=1

\documentclass[useAMS,usenatbib,usegraphicx]{mn2e}
\usepackage{color}
\usepackage{aas_macros}
\usepackage{amsmath}
\usepackage{newpxmath}
\usepackage{graphicx}
 \usepackage{multirow}
 \usepackage{float}
\usepackage{subfig}
\title[]{The relative impact of photoionizing radiation and stellar winds on different environments}
\author[S. Haid]{S. Haid$^{1}$\thanks{E-mail:
haid@ph1.uni-koeln.de}, S. Walch$^{1}$, D. Seifried$^{1}$, R. W\"unsch$^{2}$, F. Dinnbier$^{1}$ and T. Naab$^{3}$\\
$^{1}$I. Physikalisches Institut, Universit\"at zu K\"oln, Z\"ulpicher-Strasse 77, 50937 Cologne, Germany\\
$^{2}$Astronomick\'{y} \'{U}stav, Akademie v\u{e}d \u{c}esky Republiky, Bo\u{c}n\'{i} II 1401, CZ-14131 Praha, Czech Republic\\
$^{3}$Max-Planck-Insitut f\"ur Astrophysik, Karl-Schwarzschild-Strasse 1, 85741 Garching, Germany}
\begin{document}

\maketitle

\label{firstpage}

\begin{abstract}
Photoionizing radiation and stellar winds from massive stars deposit energy and momentum into the interstellar medium (ISM). They might disperse the local ISM, change its turbulent multi-phase structure, and even regulate star formation. Ionizing radiation dominates the massive stars' energy output, but the relative effect of winds might change with stellar mass and the properties of the ambient ISM. We present simulations of the interaction of stellar winds and ionizing radiation of 12, 23, and 60 M$_{\odot}$ stars within a cold neutral (CNM, $n_{0}$ = 100 cm$^{-3}$), warm neutral (WNM, $n_{0}$ = 1, 10 cm$^{-3}$) or warm ionized (WIM, $n_{0}$ = 0.1 cm$^{-3}$) medium. The \textsc{FLASH} simulations adopt the novel tree-based radiation transfer algorithm \textsc{TreeRay}. With the On-the-Spot approximation and a temperature-dependent recombination coefficient, it is coupled to a chemical network with radiative heating and cooling. In the homogeneous CNM, the total momentum injection ranges from 1.6$\times$10$^{4}$  to 4$\times$10$^{5}$ M$_{\odot}$ km s$^{-1}$  and is always dominated by the expansion of the ionized H$_{\text{II}}$ region. In the WIM, stellar winds dominate (2$\times$10$^{2}$ to 5$\times$10$^{3}$ M$_{\odot}$ km s$^{-1}$), while the input from radiation is small ($\sim$ 10$^{2}$ M$_{\odot}$ km s$^{-1}$). The WNM ($n_{0}$ = 1 cm$^{-3}$) is a transition regime. Energetically, stellar winds couple more efficiently to the ISM ($\sim$ 0.1 percent of wind luminosity) than radiation ($<$ 0.001 percent of ionizing luminosity). For estimating the impact of massive stars, the strongly mass-dependent ratios of wind to ionizing luminosity and the properties of the ambient medium have to be considered.

\end{abstract}


\section{Introduction}
\label{sec-introduction}
Feedback from massive stars in the form of ionizing radiation, stellar winds, and supernova (SN) explosions modifies the density distribution, changes the chemical composition and influences the energy budget of the environment. For young, massive stars, which are still embedded in a gravitationally collapsing cloud, these processes can counteract the contraction and prevent further accretion of material onto the star. Hence, feedback by stellar winds and ionizing radiation provides one feasible way to locally suppress star formation by dispersing the cold gas in the molecular cloud out of which the massive star has been born \citep{whitworth79, krumholz06, krumholz09c, walch12, dale15}. As a result, the SN at the end of the stars' lifetime explodes into the pre-blown bubble, which is already hot and ionized \citep{walch15a}. On the other hand, stellar feedback can also trigger star formation in the surrounding cloud at distances of several parsec from the massive star \citep{elmegreen77, whitworth79, krumholz06, gritschneder09, krumholz09c, gritschneder10, walch12, walch13, dale15}. 

On scales larger than individual molecular clouds or cloud cores, the impact of persistent stellar energy emission is still highly debated \citep{ostriker10, dobbs14, hopkins14, krumholz14, naab17}. It is likely that a more detailed understanding of the local interaction of stellar winds and ionizing radiation from massive stars, in addition to SNe, are the key to answering some of the major questions in star formation and galaxy evolution \citep{naab17}, e.g. galactic outflows might be driven by stellar feedback processes.

Followed by the first theoretical model of the effect of ionizing radiation \citep{stromgren39}, the description of the expansion of an H$_{\text{II}}$ region into a homogeneous medium has been derived by \citet{spitzer78} and extended to account for the inertia of the shell by \citet{hosokawa06}. Many modern codes have tested these analytic expressions (\citealt{bisbas15} and references therein). Recent three-dimensional simulations modify the ambient density distribution to be fractal \citep{walch12} or include dense self-gravitating objects \citep{matzner02, dale05}. Depending on the escape velocity \citep{dale12}, an embedded ionizing source might be able to disrupt a cloud or not \citep{howard17, geen15}. 

Wind-blown bubbles were first analytically discussed by \citet{castor75} and \citet{weaver77} for homogeneous media. Later, power-law environmental density distributions were studied \citep{franco90, koo92, garciasegura95a, pittard01}. The complexity of numerical simulations increased with even more realistic ambient media and the interaction with other feedback processes \citep{falle75a,garciasegura95b, garciasegura96, arthur07, dwarkadas07, toala11, rogers13}. 

However, the relative impact of stellar winds with respect to ionizing radiation is still highly debated. Judging from the amount of energy provided by the star, the first process should be negligible \citep{matzner02}. However, the conversion of radiative energy to kinetic energy is highly inefficient \citep{walch12} and thus, both processes could be important. 

Analytic estimates \citep{dyson80} and two-dimensional simulations \citep{freyer03, kroeger06, hensler08} indicate that stellar winds couple efficiently with the environmental gas, which means that a significant fraction of the wind input energy is received by the environmental gas in form of thermal and kinetic energy. In addition, the wind momentum input is fully retained. However, the impact of the wind on a surrounding molecular cloud could still be small \citep{mackey13, dale14, geen15}. The wind of a single B-star in the presence of a self-gravitating cloud is surely too weak to counteract the gravitational collapse \citep{geen15}. In simulations with Smoothed Particle Hydrodynamics, momentum-driven winds from a massive, 30 M$_{\odot}$, O-star also show little destructive behaviour \citep{dale08} but compress gas in a shell. If the shell becomes unstable or the wind expands into a medium with turbulent sub-structures, then the gas with low (lower than average) density is channelled into rarefied regions where it can leak out of the cloud \citep{harperclark09, dale13, rogers13, rosen14}. These "paths of least resistance" allow for the following SN to disperse the cloud \citep{pittard13, wareing17}. 

Studies of the impact of stellar feedback processes generally consider that the sources are embedded in dense molecular clouds where stars are born (e.g. \citealt{freyer06, arthur11, dale12, walch12, geen15, ngoumou15}). However, the environment of a massive star is likely to change rapidly. It might become warm and ionized (e.g. \citealt{felli84}) due to previously born stars in the same cluster. In addition, stars are not static. About 20-40 percent of O stars \citep{gies86, stone91} are estimated to be runaway stars, migrating at typical velocities of tens km/s up to several hundreds of km/s into the warm ionised or hot medium \citep{hoogerwerf00, huthoff02}. Runaway stars reach distances of several hundreds of parsecs from their birthplaces in a few Myr. This motivates us to study the interaction of ionising radiation and stellar winds not only with the cold dense phase of the ISM, but also with the more rarefied warm ionised phase.

In this paper, we address the question with which efficiencies stellar winds and ionizing radiation couple the provided energy and momentum to the environment. We investigate different ambient media, ranging from the prototypical cold neutral medium to a warm ionized medium. In addition, we present the first application of the novel, three-dimensional, tree-based radiative transfer method \textsc{TreeRay}. We use a single energy band to treat ionizing radiation and couple it to the employed chemical network. The network follows the evolution of 7 species (H$_{2}$, H, H$^{+}$, CO, C$^{+}$, O, e$^{-}$). We are able to self-consistently treat heating and cooling of the ambient gas.

In section \ref{sec-numericalmethod}, we briefly describe the simulation code \textsc{FLASH} 4 and the simulation setup. We also introduce the novel radiative transfer method \textsc{TreeRay} and how it is coupled to the chemical network, and compare the resulting temperatur of the different H$_{\text{II}}$ regions with the ionizing radiation Monte-Carlo code \textsc{MOCASSIN}. In section \ref{sec-combi}, we discuss the impact of the combination of stellar winds and ionizing radiation and in section \ref{sec-individual}, we show the individual and relative impact of both processes. In section \ref{sec-efficiencies}, we take a look at energy coupling efficiencies and the implications on the emission of radiative recombination cooling and X-rays. We summarize in section \ref{sec-summary}.

\section{Numerical method}
\label{sec-numericalmethod}
We use the Eulerian, adaptive mesh refinement, magneto-hydrodynamics (MHD) code \textsc{FLASH} 4 \citep{fryxell00,dubey08} with the directionally split, Bouchut HLL5R solver \citep{bouchut07, bouchut10, waagan09, waagan11}. In addition, self-gravity, radiative transfer, radiative cooling and heating (from the ionizing radiation as well as from a constant background interstellar radiation field), shielding of molecular hydrogen and CO \citep{wunsch17} and a chemical network is included (\citealt{glover07a, glover07b,  glover10}, for the implementation into \textsc{FLASH} see \citealt{walch15}). We use a stellar evolution model with a momentum-driven wind \citep{gatto17} and chemistry-coupled ionizing radiation. In this work, we do not treat magnetic fields. In the next subsections we will describe the implementations in more detail. 

\subsection{Stellar winds}
\label{subsec-wind}
To simulate the impact of stellar winds we partly follow the procedure of \citet{gatto17}. The evolution of massive stars (possible masses of 9 to 120 M$_{\odot}$) is modelled using the Geneva stellar evolution tracks from the zero-age main sequence to the Wolf-Rayet phase \citep{ekstroem12}. An initial proto-stellar phase is not included.

\begin{figure}
\includegraphics[width=0.5\textwidth]{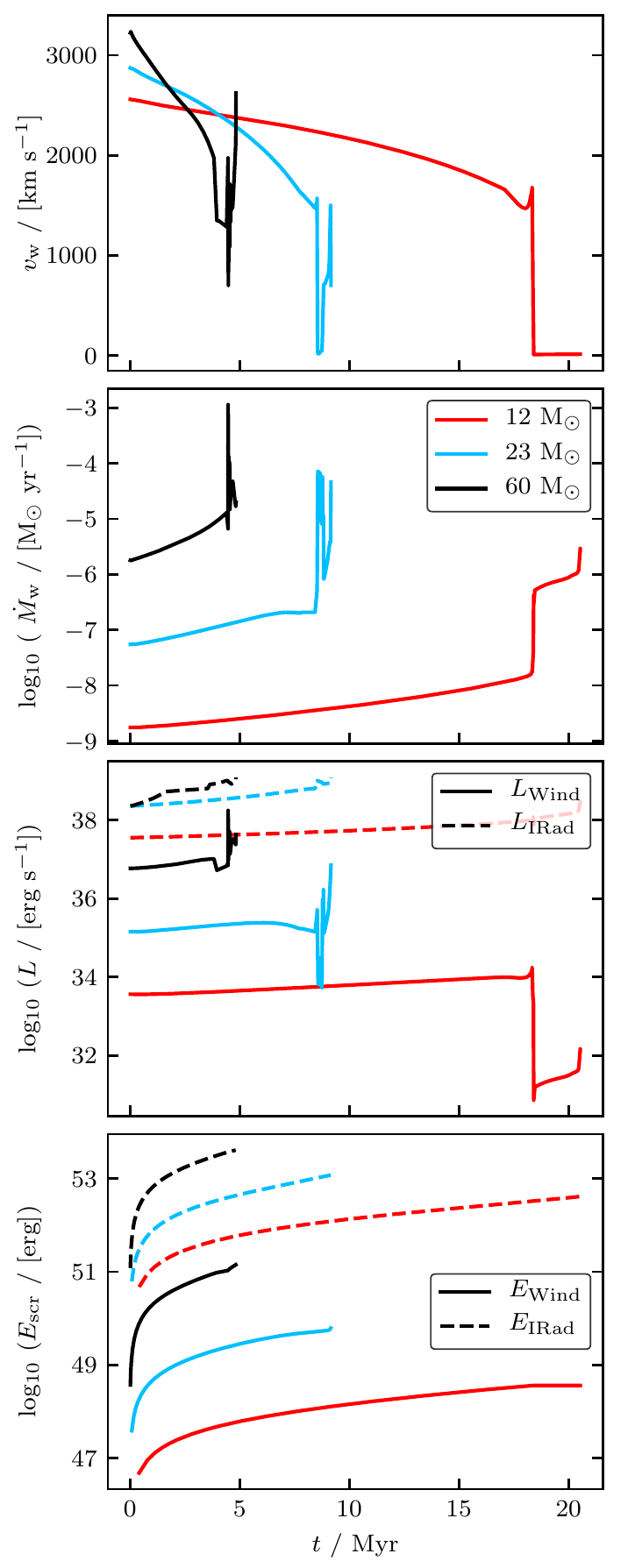} 
\caption{Wind velocities (top panel), wind mass-loss rates (second panel) as well as luminosities (third panel) and energies (bottom panel) of stellar winds (solid) and radiation (dashed) of stars with $M_{*}$ = 12 (red), 23 (blue), and 60 M$_{\odot}$ (black).}
\label{fig-evolution-tracks}
\end{figure}

The wind mass-loss rates $\dot{M}_{\text{w}}$ are taken from the tracks by \citet{ekstroem12}. The corresponding terminal wind velocities $v_{\text{w}}$ are estimated according to the evolutionary status (\citealt{puls09}, in Section 2.4 in \citealt{gatto17} and the references therein). Fig. \ref{fig-evolution-tracks} shows the time evolution of $v_{\text{w}}$ (top panel) and $\dot{M}_{\text{w}}$ (second panel) of stars with initial masses $M_{*}$ = 12, 23, and 60 M$_{\odot}$. The radiative luminosities, $L_{\text{IRad}}$, (dashed) and mechanical luminosities, $L_{\text{Wind}} = 0.5 \dot{M_{\text{w}}} v_{\text{w}}^{2}$, (solid) are shown in the third panel. The cumulative energy inputs from the sources ($E_{\text{src}}(t) = \int^{t}_{0} L\ \text{d}t$, with $L = L_{\text{IRad}}$ or  $L = L_{\text{Wind}}$) are provided in the bottom panel. 

\subsubsection{Wind implementation}
The wind model, as implemented in \textsc{FLASH} 4, is based on the injection of momentum in a given spherical volume defined by the {\it injection radius} (see section \ref{sec-simulation-setup}). In the reference frame of the star, the wind momentum points radially outward and is assumed to have a spherically symmetric distribution. Therefore, for a given $v_{\text{w}}$, the wind velocity vector in the reference frame of the star, $\boldsymbol{v}_{\text{w}}^{\text{s}}$, that is added to every cell within the injection radius is 

\begin{equation}
\boldsymbol{v}_{\text{w}}^{\text{s}} = v_{\text{w}} \frac{\boldsymbol{x} - \boldsymbol{x_{\text{s}}}}{\Vert \boldsymbol{x} - \boldsymbol{x_{\text{s}}} \Vert}
\end{equation}  
where $\mathbf{x}_{\text{s}}$ is the position of the star. We apply a Galilei transformation to obtain the wind velocity in the rest frame as
\begin{equation}
\boldsymbol{v}_{\text{w}} = \boldsymbol{v}_{\text{w}}^{\text{s}} +\boldsymbol{v}_{\text{s}}
\end{equation}  
where $\boldsymbol{v}_{\text{s}}$ is the velocity of the star. 

The total momentum $\boldsymbol{p}'$ in a cell after the wind is injected is
\begin{equation}
 \boldsymbol{p}' = m'\boldsymbol{v}' = m_{\text{g}}\boldsymbol{v}_{\text{g}} + m_{\text{w}}\boldsymbol{v_{\text{w}}} 
\end{equation}
where $m'\boldsymbol{v}'$ is the cell mass times the cell velocity after the wind injection and $m_{\text{g}}\boldsymbol{v}_{\text{g}}$ the initial momentum of the gas in the cell. The wind momentum to be injected is $m_{\text{w}}\boldsymbol{v_{\text{w}}}$ with $m_{\text{w}} = \dot{M}_{\text{w}}(t) \Delta t (\Delta x)^{3} V_{\text{inj}}^{-1}$ where $\Delta t$ is the time step, $\Delta x$ the cell size, and $V_{\text{inj}}$ the volume into which the wind is injected. 

Since we inject the mass lost by the massive star, $m_{\text{w}}$, we have to increase the internal energy\footnote{Note that the specific internal energy is not changed since $m_{\text{w}}$ has been added to the cell.} in every cell of the injection region. To compute the required internal energy input, we make the ansatz
\begin{equation}\label{eq_energy}
E_{\text{kin}}'+E_{\text{int}}' = \left(E_{\text{kin,g}} + \Delta E_{\text{kin,w}}\right) + \left(E_{\text{int,g}} + \Delta E_{\text{int,w}} \right),
\end{equation}
where the left-hand side corresponds to the total energy after the wind injection. The right-hand side is composed of the kinetic and internal energies of the gas, $E_{\text{kin,g}}$  and $E_{\text{int,g}}$, and the kinetic and internal energy of the wind, $\Delta E_{\text{kin,w}}$ and $\Delta E_{\text{int,w}}$. Since we only inject momentum, $\Delta E_{\text{int,w}} = 0$ by definition. With $E_{\text{kin}}' = \frac{ \boldsymbol{p}'^{2}}{2m'}$ and
\begin{equation}
\Delta E_{\text{kin,w}} = E_{\text{kin}}' - E_{\text{kin,g}} = \frac{ \boldsymbol{p}'^{2}}{2m'} - \frac{ \left(m_{\text{g}}\boldsymbol{v}_{\text{g}} \right)^{2}}{2m_{\text{g}}},
\end{equation}
we can solve Eq. \ref{eq_energy} for $E_{\text{int}}'$ and obtain
\begin{equation}
E_{\text{int}}' = E_{\text{int,g}} + \frac{1}{2}\frac{m_{\text{g}}m_{\text{w}}}{m_{\text{g}}+m_{\text{w}}}\left( \boldsymbol{v}_{\text{g}} - \boldsymbol{v_{\text{w}}}  \right)^{2}.
\end{equation}

%%%%%%%%%
\subsubsection{Analytic solution}
\label{sec-wind-analytic}
In a uniform medium, the momentum from the adiabatic, thin-shell evolution of stellar winds can be calculated analytically \citep{weaver77, garciasegura95a, garciasegura95b, everett10b}. The time evolution of the shock radius is \citep{weaver77, pittard13}
\begin{equation}
R_{\text{Wind}} =  \left( \frac{125}{154\pi} \right)^{0.2} \left(\frac{0.5\dot{M}_{\text{w}}  v_{\text{w}}^{2}}{\rho_{0}}\right)^{0.2}t^{0.6}.
\label{eq-weaver}
\end{equation}
%= 0.88 \left(\frac{0.5\dot{M}_{\text{w}}  v_{\text{w}}^{2}}{\rho_{0}}\right)^{0.2}t^{0.6}.
The resulting momentum input $p_{_{\text{theo, Wind}}}$ is
\begin{equation}
\label{eq-windmom}
p_{_{\text{theo, Wind}}} =  \frac{4\pi}{3}\rho_{0} 0.6 \frac{R_{\text{Wind}}^{4}}{t}, 
\end{equation}
where $\rho_{0}$ is the density of the uniform, ambient medium \citep{krumholz09b}.

\subsection{Ionizing radiation and radiative heating}
\label{subsec-radiation}

\subsubsection{Analytic solution}
\label{sec-radiation-analytic}

For the analytic treatment of ionizing radiation we consider all Lyman continuum photons with an energy of $h\bar{\nu} \geq 13.6\ \text{eV}$, which can ionize hydrogen immediately. In ionization-recombination equilibrium, the result is an H$_{\text{II}}$ region which extends to the Str\"omgren radius $R_{\text{St}}$ with
\begin{equation}
R_{\text{St}} = \left( \frac{3}{4\pi} \frac{ \dot{N}_{_{\text{LC}}}m_{\text{p}}^{2}}{\alpha_{\text{B}}\rho_{0}^{2}}\right)^{1/3},
\label{eq-spitzer1}
\end{equation}
where $\dot{N}_{_{\text{LC}}}$ is the number of emitted Lyman continuum photons per second, $m_{\text{p}}$ the proton mass, $\rho_{0}$ the ambient density, and $\alpha_{\text{B}} $ the radiative recombination rate of hydrogen to all levels above the ground state (case B recombination) 
\begin{equation}
\label{eq-alphab}
\alpha_{\text{B}} = 2.56\times10^{-13} \left(\frac{T}{10^{4}\ \text{K}}\right)^{-0.83} \left[\text{cm}^{3}\ \text{s}^{-1}\right],
\end{equation} in the range of $T$ = [5000, 20000] K \citep{tielens05}.

The temperature inside the H$_{\text{II}}$ region is immediately increased due to photoionization heating. A pressure gradient establishes at $R_{\text{St}}$ and drives a shock with the shock velocity
\begin{equation}
v_{\text{S}} = c_{\text{i}} \sqrt{\frac{4}{3}\frac{R_{\text{St}}^{1.5}}{R_{\text{IRad}}^{1.5}} - \frac{\mu_{\text{i}}T_{\text{o}}}{2\mu_{\text{o}}T_{\text{i}}}}
\label{eq-spitzer3}
\end{equation}
where  $T_{\text{0}}$ and $\mu_{\text{0}}$ are the ambient temperature and mean molecular weight and  $c_{\text{i}}$, $T_{\text{i}}$, and $\mu_{\text{i}}$ are the isothermal sound speed, the temperature, and the mean molecular weight of the ionized medium. Under the assumption that $v_{\text{S}}$ is larger than the ambient sound speed, $R_{\text{IRad}}$ is the shock radius which evolves with the Hosokawa-Inutsuka modification of the analytic Spitzer solution as \citep{spitzer78, hosokawa06, bisbas15}
\begin{equation}
R_{\text{IRad}} = R_{\text{St}}\left( 1 + \frac{7}{4} \sqrt{\frac{4}{3}} \frac{c_{\text{i}}t}{R_{\text{St}}}\right)^{4/7}.
\label{eq-spitzer2}
\end{equation}
The resulting momentum is  
\begin{equation}
\label{eq-spitzermom}
p_{_{\text{IRad}}} = \frac{4}{3}\pi \left(R_{\text{IRad}}^{3} - R_{\text{St}}^{3} \right) \rho_{0}v_{\text{S}}.
\end{equation}

\subsubsection{\textsc{TreeRay}}
\label{sec-treeray}
The transfer of ionizing radiation is calculated by a new module for the \textsc{FLASH} code called \textsc{TreeRay}. It is an extension of the \textsc{FLASH} tree solver for calculating self-gravity and the optical depth in every cell of the computational domain as described in \citet{wunsch17}. Here, we only give a basic information about \textsc{TreeRay}. A detailed description alongside with accuracy and performance tests will be presented in W\"unsch et al. (in prep).

\textsc{TreeRay} uses the octal-tree data structure constructed and updated in each time step by the tree solver, and shares it with the Gravity and Optical-Depth modules. Each node of the octal-tree represents a cuboidal collection of grid cells and stores the total gas mass contained in it, masses of individual chemical species, and the position of the mass centre. In addition to that, \textsc{TreeRay} stores for each node the total amount of the radiation luminosity generated inside the node, radiation energy flux passing through the node, and the node volume.

Before the tree is traversed for each grid cell (called {\em target cell}), a system of rays pointing from the target cell to all directions is constructed. The directions are determined by the \textsc{HEALPIX} algorithm \citep{gorski05}, which tessellates the unit sphere into elements of equal spatial angle. Each ray is then divided into segments with lengths increasing linearly with the distance from the target cell. In this way, the segment lengths correspond approximately to sizes of nodes interacting with the target cell during the tree walk if the Barnes-Hut criterion for node acceptance is used. When the tree is traversed, node densities, radiation luminosities and energy fluxes are mapped onto ray segments according to a degree of the intersection of the node volume and the volume belonging to the ray segment.

Finally, after the tree walk, the one-dimensional radiative transport equation is solved along each ray. In this work, this equation has a form corresponding to the physical processes and approximations (On-the-Spot) used in Sec. \ref{sec-radiation-analytic}, i.e. the absorption coefficient is proportional to $\alpha_{\rm B}(T) \rho^2$ and the emission coefficient is proportional to $\dot{N}_{_{\rm LC}}$ for a given source. As the radiation flux passing through a given segment from different directions has to be taken into account, the solution has to be searched for iteratively, repeating the whole process of tree construction, tree walk and radiation transport equation solving until a desired accuracy is reached. Fortunately, the solution from the previous hydrodynamic time-step can be used, and as the radiation field typically changes only slightly between times-steps, in most cases only one or two iterations in each time step are needed\footnote{Note that the {\sc FLASH} code uses global time steps and that the time step is always limited by the CFL condition of the fast stellar wind, which is much more restrictive than the progress of the D-type ionization front.}.

For the performed simulations, we use 48 rays and the tree solver with the Barnes-Hut acceptance criterion with limit angle $\theta_{\rm lim} = 0.5$. The code is benchmarked for the expanding H$_{\text{II}}$ region \citep{bisbas15}. In this paper, we extend the prescription with a temperature-dependent recombination coefficient and couple \textsc{TreeRay}  to the chemical network.

The main advantage of \textsc{TreeRay} is that the computational cost is basically independent of the number of sources. Therefore, it can be readily used to simulate the radiative feedback of many stars in e.g. clusters in full three-dimensional MHD simulations.

\subsubsection{Ionizing radiation heating}
We assume that all sources emit a black-body spectrum with an effective stellar temperature $T_{*}$ given by the aforementioned stellar tracks. Thus, the mean ionizing photon energy $h\bar{\nu}$ is
\begin{equation}
h \bar{\nu} = \frac{\int_{\nu_{\text{T}}}^{\infty} B_{\nu}\text{d}\nu}{\int_{\nu_{\text{T}}}^{\infty}\frac{B_{\nu}}{h\nu}\ \text{d}\nu} = \frac{F}{F_{\text{ph}}}
\end{equation}
where $h$ is the Planck constant, $\nu_{\text{T}}$ = 13.6 eV $h^{-1}$ is the threshold frequency for hydrogen ionization, $B_{\nu} = B_{\nu}(T_{*})$ is the Planck function, $F$ the energy flux and $F_{\text{ph}}$ the photon flux \citep{rybicki04}. Both fluxes are provided by \textsc{TreeRay} for every cell in the computational domain. 

The heating rate $\Gamma_{\text{ih}}$ in ionization-recombination equilibrium is calculated with \citep{tielens05}
\begin{equation}
\Gamma_{\text{ih}} = F_{\text{ph}} \sigma E_{\bar{\nu} - \nu_{T}} = n_{\text{H}}^{2} \alpha_{\text{B}}  h \left(\bar{\nu} - \nu_{\text{T}} \right) 
\label{eq-heatingrate}
\end{equation}
where $E_{\bar{\nu} - \nu_{T}}=h \left(\bar{\nu} - \nu_{\text{T}} \right)$ is the average excess energy of the ionizing photons, $\sigma$ is the hydrogen photoionization cross-section, $n_{\text{H}}$ the hydrogen number density, and $\alpha_{\text{B}}$ (see Eq. \ref{eq-alphab}) the radiative recombination rate.

The heating rate and number of ionizing photons are provided to the chemistry module (see Section \ref{sec-chemistry}). There, the temperature will be increased self-consistently by balancing heating and cooling processes and the hydrogen species will be updated using the given photoionization rate.

\subsection{Gas cooling, heating and chemistry}
\label{sec-chemistry}

We include a simple chemical network, which is explained in detail in \cite{walch15}. It is based on \citet{glover07b, glover07a, glover10} and \citet{nelson97} to follow the abundances of seven chemical species: molecular, atomic and ionized hydrogen as well as carbon monoxide, ionized carbon, atomic oxygen and free electrons (H$_{2}$, H, H$^{+}$, CO, C$^{+}$, O, e$^{-}$). The gas has solar metallicity \citep{sembach00} with fixed elemental abundances of carbon, oxygen and silicon ($x_{\text{C}}$ = 1.4 $\times$ 10$^{-4}$, $x_{\text{O}}$ = 3.16 $\times$ 10$^{-4}$, $x_{\text{Si}}$ = 1.5 $\times$ 10$^{-5}$) and the dust-to-gas mass ratio is set to 0.01. We include a background interstellar radiation field (ISRF) of homogeneous strength G$_{0}$ = 1.7 \citep{habing68, draine78}. To assume the ISRF to be constant near a massive star is an approximation. However, even a 100 times higher radiation strength increases the temperature in the medium by only 12 percent (see Section \ref{sec-G0} in the Appendix). Thus the ambient pressure counteracting the expanding shock would change only marginally. For this reason, we here only focus on the case of G$_{0}$ = 1.7. The ISRF is attenuated in shielded regions depending on the column densities of total gas, H$_{2}$, and CO. Thus, we consider dust shielding and molecular (self-)shielding for H$_{2}$ and CO \citep{glover10, walch15} by calculating the shielding coefficients with the \textsc{TreeRay} Optical-Depth module, described and tested in \citep{wunsch17}.

For gas with temperatures above $\sim$ 10$^4$ K we model the cooling rates according to \citet{gnat12} in collisional ionization equilibrium. Non-equilibrium cooling for the respective species is applied at lower temperatures (also for Lyman $\alpha$). Within the H$_{\text{II}}$ region, we neglect both C$^{+}$ and O cooling because these species are assumed to be in a higher ionization state.

Heating rates include the photoelectric effect, cosmic ray ionization with a rate of $\xi$ = 3$\times$10$^{-17}$ s$^{-1}$, and X-ray ionization by \cite{wolfire95b}. In this work we additionally include the heating by photoionization from the central star (see Eq. \ref{eq-heatingrate}). 

Note that, since we only consider the radiative transfer in a single energy band (all photons in the Lyman continuum), we do not distinguish between the direct ionization of H and H$_{2}$, as necessary for detailed models of photon-dominated regions \citep{rollig07}. However, photon-dominated regions in an early evolutionary stage are considered thin and unresolved in three-dimensional simulations of feedback in MCs (see Eq. 1 in \citealt{bisbas15}). During the evolution this region will widen, however the treatment of this late stage is beyond the scope of our simulations.

\subsection{Simulation setup}
\label{sec-simulation-setup}
We use cubic boxes with a side length of 51 pc. The generic grid resolution is 0.4 pc with a maximum resolution of 0.2 pc refining on the source and the density fluctuations of the shell. The computational domain is homogeneously filled with initially warm and ionized gas (WIM) or with cold, predominantly neutral gas (CNM). The initial densities are $\rho_{0}= 2.1\times10^{-25}$ g cm$^{-3}$ and $2.1\times10^{-22}$ g cm$^{-3}$ and temperatures $T_{0}=10^4$ K and 20 K, respectively. The according number densities for an assumed mean molecular weight\footnote{Although the mean molecular weight is computed self-consistently using the {\sc FLASH} Multispecies module, and thus the resulting number densities are not exactly equal to 0.1 and 100 cm$^{-3}$ we refer rather to $n_{0}$ than $\rho_{0}$ throughout most of the paper.} of 1.4 are $n_{0} = 0.1$ and 100 cm$^{-3}$. 

The chemical species are initialized using fractional abundances. In the WIM, the initial H$^{+}$ abundance $n_{\text{H+}}/n_{\text{H, tot}}$ = 0.98, and the other 2 percent are neutral. In the CNM, we initialize $n_{\text{H}}/n_{\text{H, tot}}$ = 0.5 and $n_{\text{H2}}/n_{\text{H, tot}}$ = 0.25. Independent of the medium, carbon is always ionized, $n_{\text{C+}}/n_{\text{C, tot}} = 1$.

We consider three different single stars with initial masses of $M_{*}$ = 12, 23, and 60 M$_{\odot}$. The star is always placed in the center of the domain and emits a stellar wind and/or ionizing radiation. The spherical wind injection region is 12 cells in radius on the highest level of refinement, corresponding to 2.4 pc. Table \ref{tab-homogeneous1} summarises the 18 simulations, which were performed for this section.

\begin{table}
\caption{We list the simulations with the employed stellar process(es), ambient medium, and mass of the central star. Abbreviations: \textit{WIM} warm ionized medium ($n_{0}$ = 0.1 cm$^{-3}$,  $T_{0}$ = 10$^{4}$ K), \textit{CNM} cold neutral medium ($n_{0}$ = 100 cm$^{-3}$, $T_{0}$ = 20 K)}
\label{tab-homogeneous1}
\centering
\begin{tabular}{|c|c|cc}
\hline 
\rule[-1ex]{0pt}{2.5ex}  Wind & IRad & Media & Sources [M$_{\odot}$] \\ 
\hline
\hline  
\rule[-1ex]{0pt}{2.5ex}   & X & WIM, CNM & 12, 23, 60 \\ 
\hline 
\rule[-1ex]{0pt}{2.5ex}  X &  & WIM, CNM & 12, 23, 60 \\
\hline 
\rule[-1ex]{0pt}{2.5ex}  X & X & WIM, CNM & 12, 23, 60 \\ 
\hline 
\end{tabular} 
\end{table}

\subsection{Benchmark with \textsc{MOCASSIN}}
\label{sec-simulation-heating}

\begin{figure}
\includegraphics[width=0.5\textwidth]{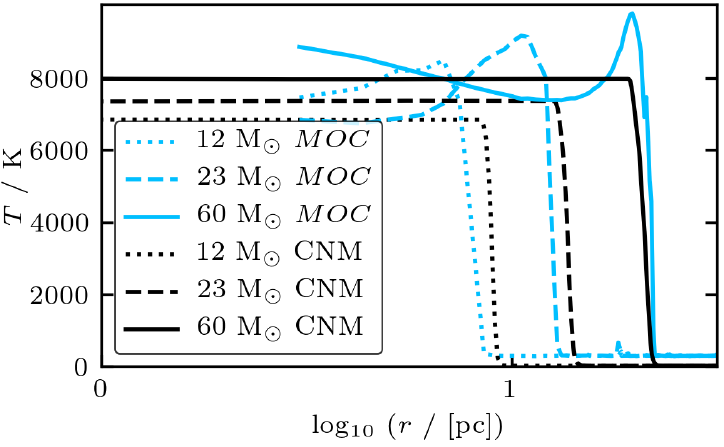} 
\caption{The radial temperature distributions obtained from the simulations of ionizing radiation in the CNM (black), which are compared to simulations with the photoionization Monte-Carlo code \textsc{MOCASSIN} (blue). The profiles are shown for stars with  $M_{*}$ = 12 (dotted), 23 (dashed), and 60 M$_{\odot}$ (solid) at $t$ = 2.5 Myr.}
\label{fig-mocassin} 
\end{figure}

First, we compare the results to the three-dimensional Monte-Carlo photoionization code \textsc{MOCASSIN} \citep{ercolano03}. For this purpose, we use three \textsc{FLASH} simulations of massive stars with $M_{*}$ = 12, 23, and 60 M$_{\odot}$, which are the source of ionizing radiation only and embedded in the CNM. As input for \textsc{MOCASSIN} we deliver the CNM conditions, the mean stellar temperature of each star, $T_{*}=2.8\times$10$^{4}$ , 3.7$\times$10$^{4}$ and 4.7$\times$10$^{4}$ K, and a constant ionizing photon rate of 2.4$\times$10$^{48}$, 3.2$\times$10$^{49}$ and 2.4$\times$10$^{50}$ s$^{-1}$ for increasing stellar masses. Both parameters are time averages from the stellar tracks over the initial period of 2.5 Myr.  

Fig. \ref{fig-mocassin} compares the radial profiles of \textsc{MOCASSIN} (blue) to the {\sc FLASH} results (black) at $t$ = 2.5 Myr. With increasing stellar mass, the relative errors of the position of the shock front are 6, 8, and 11 percent. The temperature structure inside the H$_{\text{II}}$ region as calculated by \textsc{MOCASSIN} cannot be reproduced by our single-energy-band model because we are not able to treat the hardening of the radiation field at increasing distance from the central star. Yet, the volume-averaged mean temperatures agree to within 8, 5, and $\sim$ 1 percent (see Table \ref{tab-moccasin}). Therefore the mean temperatures are representative.

For the WIM, we obtain a constant radial temperature profile inside the computational domain. This is due to the fact that the Str\"omgren radius lies at a few 100 pc (see Eq. \ref{eq-spitzer1}). It is impossible that an expanding shock establishes as the pressure gradient over the shell in this medium is negligible.

\begin{table}
\caption{Comparison of the mean H$_{\text{II}}$ region temperatures with $M_{*}$ = 12, 23, and 60 M$_{\odot}$ in WIM and CNM. The second and third columns are values obtained from the code \textsc{MOCCASIN}. The last two columns show the mean temperatures inside the H$_{\text{II}}$ regions from the {\sc FLASH} simulations.}
\label{tab-moccasin}
\centering
\begin{tabular}{|c|c|c||c|c}
\hline 
\rule[-1ex]{0pt}{2.5ex}  &\multicolumn{2}{c}{\textsc{MOCCASIN}} & \multicolumn{2}{c}{{\sc FLASH}} \\ 
\rule[-1ex]{0pt}{2.5ex} Sources  & $\bar{T}_{\text{WIM}}$ [K] & $\bar{T}_{\text{CNM}}$  [K] & $\bar{T}_{\text{WIM}}$  [K]& $\bar{T}_{\text{CNM}}$  [K]\\ 
\hline
\hline  
\rule[-1ex]{0pt}{2.5ex} 12 M$_{\odot}$ & 7190 & 7730 & 7190 & 7160 \\ 
\hline 
\rule[-1ex]{0pt}{2.5ex} 23 M$_{\odot}$ & 7760 & 7710 & 7700 & 7340 \\
\hline 
\rule[-1ex]{0pt}{2.5ex} 60 M$_{\odot}$ & 8220 & 7990 & 8150 & 7940 \\ 
\hline 
\end{tabular} 
\end{table}

\section{Combined impact of stellar wind and ionizing radiation}
\label{sec-combi}

\begin{figure*}
\includegraphics[width=\textwidth]{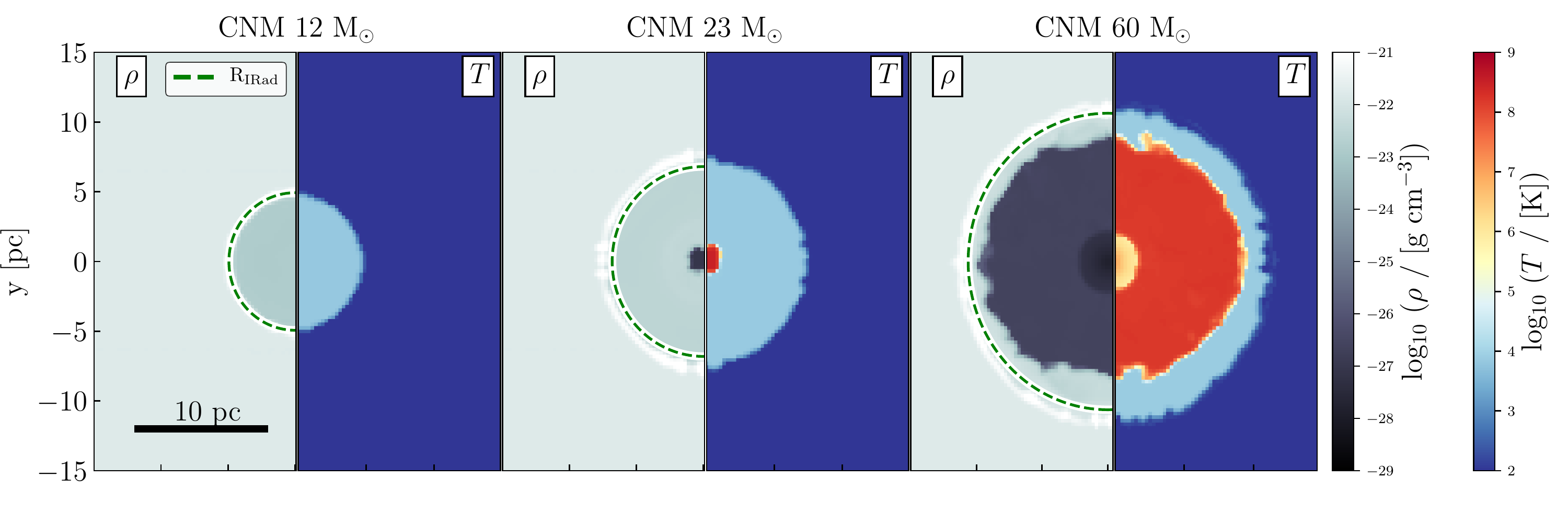} 
\includegraphics[width=\textwidth]{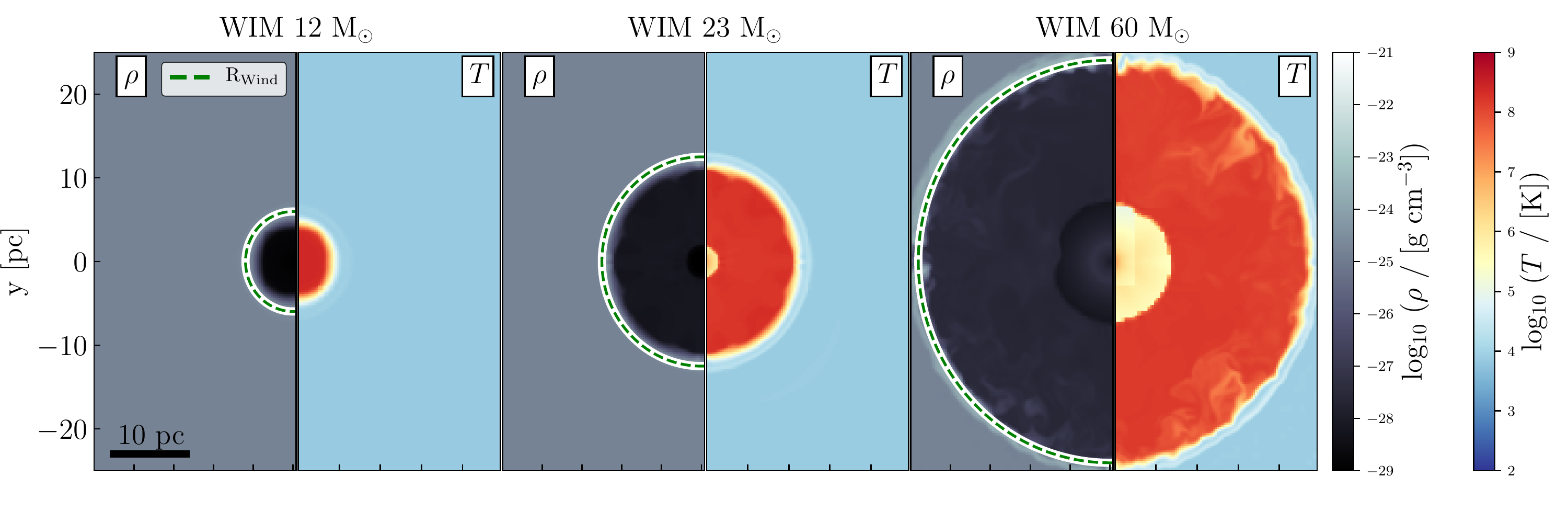} 
\caption{Effect of ionizing radiation and stellar wind with $M_{*}$ = 12, 23, and 60 M$_{\odot}$ (from left to right)  after 0.76 Myr in the CNM ($n_{0}$ = 100 cm$^{-3}$, $T_{0}$ = 20 K, top) and after 0.2 Myr in the WIM ($n_{0}$ = 0.1 cm$^{-3}$, $T_{0}$ = 10$^{4}$ K, bottom). Shown are slices through the $z=0$ plane in density (left subpanel) and temperature (right subpanel) for each source. The simulations in the same row share the same length and color scale. The top row is a zoom of the total computational domain with a length scale of 30 pc. The green lines indicate the theoretical radiation-driven shock radius (top row; Eq. \ref{eq-spitzer2}) and the analytic wind-driven shock radius (bottom row; Eq. \ref{eq-weaver}).}
\label{fig-comparison} 
\end{figure*}

Fig. \ref{fig-comparison} demonstrates the impact of the combination of both, stellar winds and ionizing radiation feedback, with $M_{*}$ = 12, 23, and 60 M$_{\odot}$ (from left to right), each in the CNM at $t$ = 0.76 Myr (top) and the WIM at $t$ = 0.2 Myr (bottom). The first time is chosen as a representative example and at the second time, the wind shock of the star with $M_{*}$ = 60 M$_{\odot}$ has reached the computational boundary. Density (left subpanels) and temperature (right subpanels) structures are shown as slices in the $z=0$ plane. Note the different length scales in the top and bottom panels. The green, dashed lines show the theoretically predicted shock radii. In the CNM, we only show these for ionizing radiation $R_{\text{IRad}}$ (Eq. \ref{eq-spitzer2}) and in the WIM only for stellar winds $R_{\text{Wind}}$ (Eq. \ref{eq-weaver}). The predicted shock radii are essentially equivalent to the computed, radially averaged shock radii, which increase with stellar mass from 5.1 pc, to 8.2 pc and 12.5 pc in the CNM and from 5.0 pc, to 11.8 pc, and 23.9 pc in the WIM for $M_{*}$ = 12, 23, and 60 M$_{\odot}$, respectively.

Inspecting Fig. \ref{fig-comparison}, we find that the gas inside the bubble has distinctive temperatures, depending on the driving process. The warm and ionized gas with a temperature of $\sim$ 8000 K is produced by ionizing radiation. The hot material with temperatures of some 10$^{7}$ K is shock-heated by stellar winds.

In the CNM (top panels), the shock is driven by ionizing radiation. The impact of stellar winds increases with the mass of the stellar source, respectively with the emitted wind luminosity (see Fig. \ref{fig-evolution-tracks}). For the star with $M_{*}$ = 12 M$_{\odot}$, the emitted wind energy is negligible compared to the emitted radiative energy. Around the star with $M_{*}$ = 23 M$_{\odot}$, the innermost $\sim$ 2 pc are shock-heated by the wind. Around the most massive star, about 80 percent of the expanding H$_{\text{II}}$ region is filled with hot but rarefied gas. Only the outer $\sim$ 4 pc are not yet affected by wind. In the centre a so-called free-wind region establishes, where the wind expands hypersonically and undisturbed.

In the case of the WIM (bottom panels), stellar winds are driving the expansion. The kinematic effect of ionizing radiation is negligible as the region, which is photoionized by the central source, is not able to expand supersonically into the warm medium. However, radiation still influences the ambient medium by preventing recombination and by providing extra heating which counteracts the cooling of the gas. Thus, the temperature remains at $\sim$ 8000 K. Without radiative support the temperature would cool down to $\sim$ 6000 K within $\sim$ 1.8 Myr.

Given the agreement of shock radii and the different theoretical predictions, as well as the fact that radiation has a stronger impact in the CNM, whereas in the WIM it is vice versa, shows that the impact of each stellar process is media-dependent.

\begin{figure}
\includegraphics[width=0.5\textwidth]{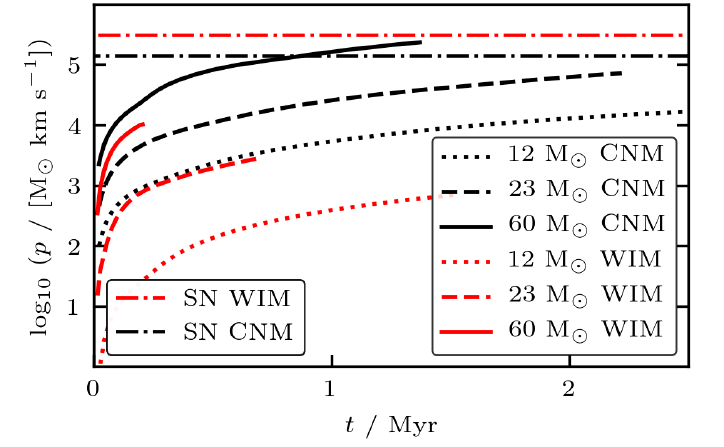} 
\caption{Evolution of the radial momentum $p$ in the CNM (black) and the WIM (red) for sources with $M_{*}$ = 12 (dotted), 23 (dashed) and 60 (solid) M$_{\odot}$  created by the combination of stellar wind and ionizing radiation. Lines terminate when the front shock has reached the boundary of the computational domain. The corresponding SN momenta in the WIM and CNM at the beginning of the momentum-conserving snowplough phase \citep{haid16} are shown as horizontal, dash-dotted lines.}
\label{fig-transferefficiency_homogeneous}
\end{figure}

In Fig. \ref{fig-transferefficiency_homogeneous} we show the time evolution of the radial momenta $p$ from the combination of stellar winds and ionizing radiation for the $M_{*}$ = 12 (dotted), 23 (dashed), and 60 (solid) M$_{\odot}$ in the CNM (black) and the WIM (red). For comparison, we show the corresponding momentum input of SNe obtained at the beginning of the momentum-conserving snowplough phase \citep[dashed-dotted lines, ][]{haid16}. This SN model assumes that the blast wave expands into a uniform ambient medium with CNM or WIM conditions and the corresponding momentum input should therefore be understood as an upper limit. Since stellar wind and ionizing radiation feedback evacuate a bubble and compress the swept-up mass in a dense shell long before the SN explosion, SN remnants might instantaneously experience significant radiative cooling when hitting the swept-up shell. This would drastically lower the final momentum input of the SN \citep{walch15a, haid16}. 

Although we consider this maximum momentum input of a type II SN, we find that the momentum input caused by a massive star with $M_{*} = 60\;{\rm M}_{\odot}$ exceeds the SN momentum input in CNM conditions with $p = 2.5\times$10$^{5}$ M$_{\odot}$ km s$^{-1}$ after only 1.5 Myr. We predict that at a later time ($\sim$ 3 Myr) also the star with $M_{*}$ = 23 M$_{\odot}$ will rise above the momentum input from a single SN. The momentum input in the WIM is systematically lower than in the CNM. The relative difference is a factor of $\sim$ 12, 6, and 2 for increasing stellar masses.

\subsection{Radial profiles of chemical abundances}
\label{sec-combi-radial}

\begin{figure*}
\includegraphics[width=1\textwidth]{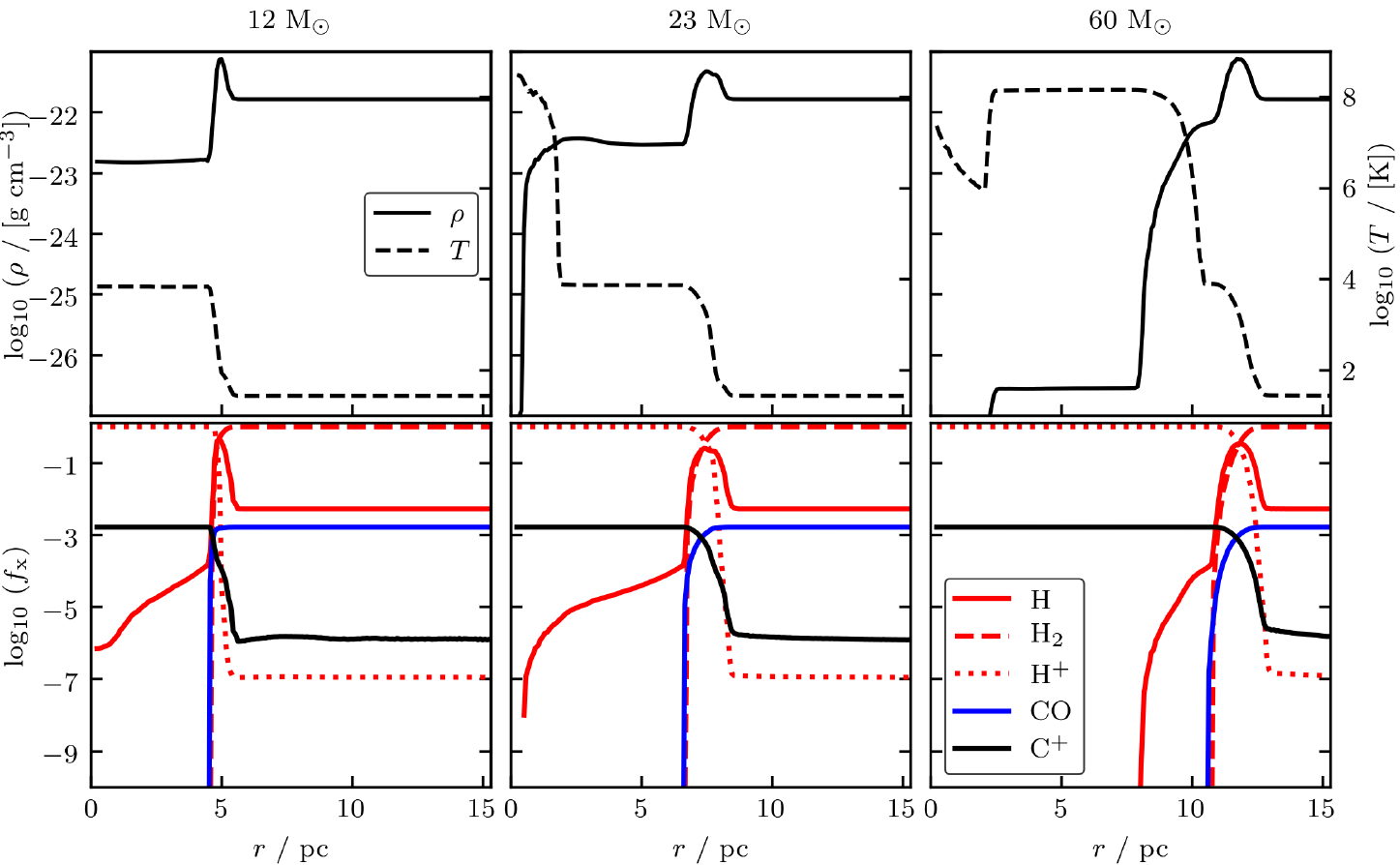} 
\caption{Radial profiles of the simulations with ionizing radiation and stellar winds for stars with $M_{*}$ = 12, 23, and 60 M$_{\odot}$ in the CNM at $t$ = 0.76 Myr. We show the radially averaged density $\rho$ (solid, left axis) and temperature $T$ (dashed, right axis) in the top panels and the mass-weighted abundances $f_{\text{x}}$ of the species H (red, solid), H$_{2}$ (red, dashed) , H$^{+}$ (red, dotted), CO (blue) and  C$^{+}$ (black) in the bottom panels.}
\label{fig-radialprofile1} 
\end{figure*}

By coupling the chemistry and the radiative transfer module, we are able to reproduce chemical transitions in shocked regions. In Fig. \ref{fig-radialprofile1} we show the radially averaged profiles around a star with $M_{*}$ = 12, 23, and 60 M$_{\odot}$ embedded in the CNM at $t$ = 0.76 Myr. Density $\rho$ (solid, left axis) and temperature $T$ (dashed, right axis) share the top panel. The bottom panel includes the mass-weighted abundance $f$ of the species H (red, solid), H$_{2}$ (red, dashed), H$^{+}$ (red, dotted), CO (blue) and  C$^{+}$ (black). The mass-weighted abundance is defined as $ f_{\text{x}} \equiv M_{\text{x}} / M_{\text{tot}}$ where $M_{\text{x}}$ is the mass of species x and $M_{\text{tot}}$ the total mass. 

The density and temperature profiles correspond to radiation-driven bubbles \citep{bisbas15} with central wind heated regions in different stages of their evolution. With increasing stellar mass the shock positions move to larger radii. 

Around the star with $M_{*}$ = 12 M$_{\odot}$, an H$_{\text{II}}$ region evolves with an average temperature and average density of $\sim$ 7200 K and 2.1$\times$ 10$^{-23}$ g cm$^{-3}$. Stellar winds show no influence. 

In the vicinity ($r< 1.9$ pc) of the star with $M_{*} = 23\;{\rm M}_{\odot}$, the wind establishes a small region filled with hot, rarefied gas ($n \sim 0.1 {\rm cm}^{-3}$, $T \sim 10^{8}$K). This compresses the gas in the H$_{\text{II}}$ region to an average density of 4.2$\times$ 10$^{-23}$ g cm$^{-3}$. The corresponding average temperature increases to $\sim$ 7300 K.

Around the star with $M_{*}$ = 60 M$_{\odot}$, the region filled with the shocked stellar wind occupies around 80 percent of the volume, which is enclosed by the ionization front. The hot but rarefied gas in the wind bubble has an average temperature of $T \sim 10^{8}$ K and an average density of $\sim 3\times10^{-2}\;{\rm cm}^{-3}$. The remaining  H$_{\text{II}}$ region is compressed into a layer of thickness $\sim$ 4 pc. There the average temperature and density is $\sim$ 8200 K and $\sim$ 8.6$\times$ 10$^{-23}$ g cm$^{-3}$. In the centre a free-wind region evolves. 

The radial profiles of the chemical abundances are qualitatively and quantitatively very similar for different M$_{*}$. Inside the H$_{\text{II}}$ region almost all hydrogen is ionized but the abundance of atomic hydrogen increases with increasing distance to the source. Ionized hydrogen drops by 6 orders of magnitude at the ionization front. The abundance of C$^{+}$ drops by 3 orders of magnitude and CO forms as the ionizing radiation becomes increasingly shielded. The outside medium contains predominantly molecular hydrogen. 

We refer to the Appendix (Appendix \ref{sec-app-radial} and Fig. \ref{fig-app-radialprofiles}) for the radial profiles of stars with $M_{*}$ = 12, 23, and 60 M$_{\odot}$ in the WIM and addition profiles of e.g. pressure and radial velocity.

\section{Relative importance of stellar wind and ionizing radiation}
\label{sec-individual}

\begin{figure*}
\includegraphics[width=\textwidth]{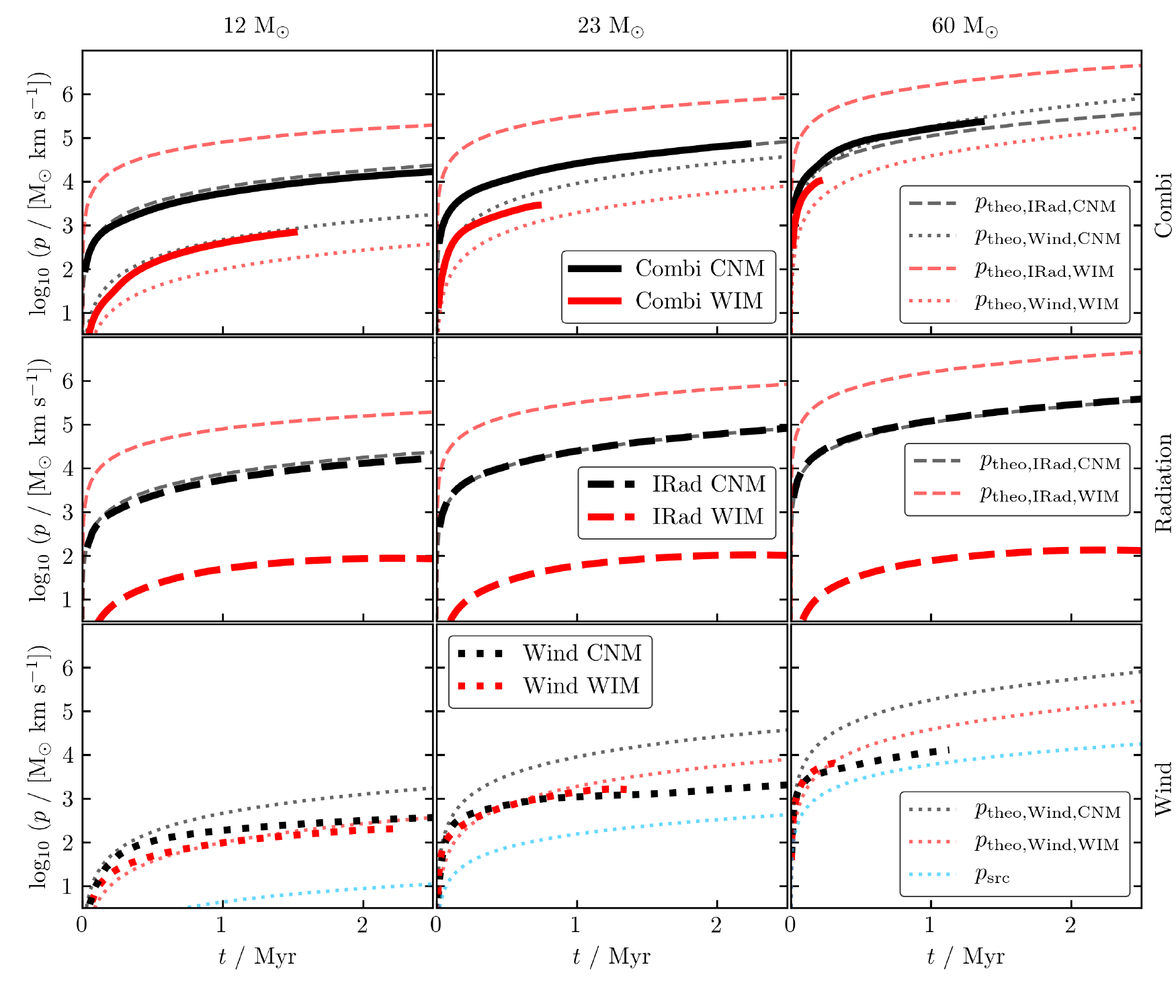} 
\caption{Evolution of the radial momentum input in the ambient medium from stars with $M_{*}$ = 12, 23, and 60 M$_{\odot}$ (from left to right) in a homogeneous WIM (red) or CNM (black) from ionizing radiation (middle row, dashed), stellar winds (bottom row, dotted), and both processes (top row, solid; see Fig. \ref{fig-transferefficiency_homogeneous}). The thin lines show the theoretical momentum input $p_{_{\text{theo}}}$ (for radiation see Eq. \ref{eq-spitzermom}, for wind see Eq. \ref{eq-windmom}). The blue lines in the bottom row indicate the wind momentum input from the source $p_{_{\text{src}}}$. In case the shown lines stop before 2.5 Myr, the feedback bubble expands out of the computational domain.}
\label{fig-comparison_sim_spi_wea}
\end{figure*}

As indicated in Section \ref{sec-combi}, the structure of the developing feedback bubble depends on the ambient medium (CNM or WIM), which suggests that the impact of stellar winds and ionizing radiation strongly depend on the medium they interact with. In order to study this more quantitatively, we carry out 12 additional simulations (see Table \ref{tab-homogeneous1}) with either wind feedback or ionizing radiation feedback. 

In Fig. \ref{fig-comparison_sim_spi_wea}, we compare the evolution of radial momenta measured in the gaseous environment of stars with $M_{*}$  = 12, 23, and 60 M$_{\odot}$ (from left to right) for simulations with either ionizing radiation (middle panel, dashed) or stellar winds (bottom panel, dotted) in the CNM (black) and the WIM (red). For comparison, the results from the combination of both feedback processes are shown in the top row (solid lines; same as Fig. \ref{fig-transferefficiency_homogeneous}). Note, that some evolutions stop before 2.5 Myr because the feedback bubble expands out of the computational domain. Therefore, when we evaluate the momentum at the end of the simulation we provide the corresponding time. 

We include the analytic estimates for ionizing radiation $p_{_{\text{theo, IRad}}}$ (thin, dashed lines, top and middle panel, Eq. \ref{eq-spitzermom}) and for stellar winds $p_{_{\text{theo, Wind}}}$ (thin, dotted lines, top and bottom panel, Eq. \ref{eq-windmom}). We also show the emitted wind momentum $p_{_{\text{src}}}(t) = \int_{0}^{t} \dot{M}_{\text{w}} v_{\text{w}}\ \text{d}t$ (blue, dotted lines) in the bottom panel. The momentum $p_{_{\text{src}}}$ is the minimum radial momentum injected into the ambient medium. Similar to an expanding SN blast wave \citep{haid16}, the feedback-driven, expanding shell gains additional radial momentum as a function of time. Therefore, the computed momenta are always larger than $p_{_{\text{src}}}$ if a stellar wind is present. For the radiation feedback, there is no minimum momentum input, because no mass-loss of the star is associated with photoionization heating. Radial outward momentum can only be generated if the radiation couples, i.e. interacts with the ambient gas. However, emitted UV radiation generates additional momentum by radiation pressure (see Appendix \ref{sec-radpres}). We want to point out, that this second process floors the minimum radiative momentum input. 

In the CNM (black), the momentum input at $t$ = 2.5 Myr from ionizing radiation (middle panels) is 1.6$\times$10$^{4}$, 8.4$\times$10$^{4}$, and 4.0$\times$10$^{5}$ M$_{\odot}$ km s$^{-1}$ with increasing stellar mass. The momentum evolution closely follows the analytic estimate and agrees with previous works by e.g. \citet{bisbas15} and \citet{geen15}. For stellar winds (bottom row) the corresponding momenta are 3.0$\times$10$^{2}$ ($t$ = 2.5 Myr), 1.9$\times$10$^{3}$ ($t$ = 2.5 Myr), and 10$^{4}$ M$_{\odot}$ km s$^{-1}$ ($t$ = 1.1 Myr). These values differ significantly from the theoretical predictions as the momentum evolution in Section \ref{subsec-wind} assumes no radiative losses. We find that the temperature in the simulated wind bubble is slightly lower and the density is slightly higher than in the adiabatic case due to cooling, where radiative cooling sets in at about 0.1 Myr. Therefore, the shock speed is lower and the shock radius lags behind $R_{_{\text{Wind}}}$, leading to a smaller swept-up mass and less radial momentum gain. The momentum gained by the combination of both feedback processes differs little from the momentum gained by ionizing radiation alone, with a relative difference of $\sim$ 1, 9, and 23 percent for increasing stellar mass. Thus, ionizing radiation is the dominant source of momentum, driving a shock in the CNM, and the contribution of stellar winds is small \citep[see also Fig. \ref{fig-comparison} and Fig. \ref{fig-transferefficiency_homogeneous}, top panel; this agrees with previous results by e.g.][]{dale08, ngoumou15}. 
%Similar numerical simulations agree with our results for ionizing radiation \citep{hosokawa06, baczynski15, bisbas15, hu17} and stellar winds \citep{dale08, ngoumou15} both in the CNM and for stars up to $M_{*}$ = 30 M$_{\odot}$.

In the WIM (red), ionizing radiation does not fully couple to the ambient medium. Therefore, it creates very little radial momentum of $\sim$ 10$^{2}$ M$_{\odot}$ km s$^{-1}$, independent of the stellar mass. The theoretical predictions disagree with the simulation results because they assume that the interior sound speed is significantly larger than the ambient sound speed. This requirement is not fulfilled in the WIM. Stellar winds generate approximative momenta between some 10$^{2}$ up to a few 10$^{3}$ M$_{\odot}$ km s$^{-1}$, which is almost the same as gained in the CNM. 

In the WIM, we find that the combination of both processes is dominated by stellar winds. Interestingly, the simulations with combined feedback (top row) follow the analytical estimates for a longer time than the simulations with wind feedback only. 

Overall, we find that stellar winds dominate in the WIM. Ionizing radiation dominates in the CNM, but is unable to expand significantly into ambient media with temperatures similar or higher to its interior. Note that the sum of momenta from individual processes is not necessarily equal to the momentum input from combined stellar feedback, $p_{_{\text{theo, IRad}}} + p_{_{\text{theo, Wind}}} \neq p_{_{\text{Combi}}}$ \citep{freyer03}. In the CNM, the feedback from both processes $p_{_{\text{Combi}}}$ is larger than $p_{_{\text{IRad}}}+ p_{_{\text{Wind}}}$ by $\sim$ 1, 3, and 23 percent with increasing stellar mass. In the WIM, the difference is a factor of $\sim$ 3.2, 2.8, and 1.9.

\subsection{The relative impact of stellar winds and ionizing radiation}
\label{sec-individual-relative}
\begin{figure}
\includegraphics[width=0.5\textwidth]{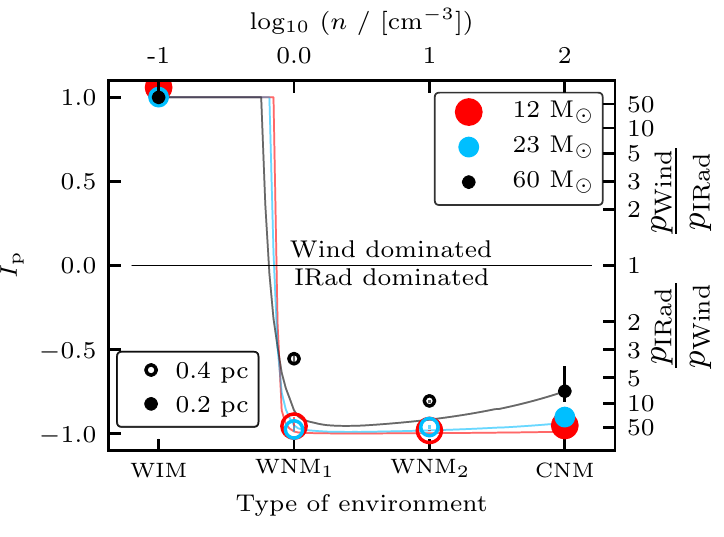} 
\caption{The relative impact of stellar winds and ionizing radiation gives an estimate which process is dominant. The horizontal line separates the wind-dominated (upper part) from the radiation-dominated (lower part). The markers show the time average relative impact $I_{\text{p}}$ of the two feedback processes. The colours represent $M_{*}$ = 12 (red), 23 (blue), and 60 M$_{\odot}$ (black). The lines show the relative impact derived from the analytic estimates. The vertical lines show the maximum and minimum values during the evolution. Full and empty markers show simulations with effective resolutions of 0.2 pc and 0.4 pc, respectively. The right ordinate shows the fraction of the dominant to the subordinate process (upper part $\frac{p_{_{\text{Wind}}}}{p_{_{\text{IRad}}}}$, lower part $\frac{p_{_{\text{IRad}}}}{p_{_{\text{Wind}}}}$). We investigate $I_{\text{p}}$ in four different environments. In warm ambient media the wind momentum input dominates, whereas the opposite applies in cold ambient media. }
\label{fig-transferefficiency_rad_wind_mom}
\end{figure} 

\begin{table}
\caption{We list the feedback processes, media and masses of the stellar sources of performed simulations with lower effective resolution of 0.4 pc. Abbreviations: \textit{WNM$_{1}$} warm neutral medium ($n_{0}$ = 1 cm$^{-3}$,  $T_{0}$ = 2$\times$10$^{3}$ K, neutral), \textit{WNM$_{2}$} warm neutral medium ($n_{0}$ = 10 cm$^{-3}$,  $T_{0}$ = 200 K, neutral)}
\centering
\label{tab-lowres}
\begin{tabular}{|c|c|cc}
\hline 
\rule[-1ex]{0pt}{1.ex} Wind & IRad & Media & Sources [M$_{\odot}$] \\ 
\hline
\hline  
\rule[-1ex]{0pt}{1.ex} & X & WNM$_{1}$, WNM$_{2}$ & 12, 23, 60 \\
\hline 
\rule[-1ex]{0pt}{1.ex} X & & WNM$_{1}$, WNM$_{2}$& 12, 23, 60 \\
\hline 
\end{tabular} 
\end{table}

In order to compare the momentum input of ionizing radiation and stellar winds, we define the relative impact $I_{\text{p}}$ as
\begin{equation}
\label{eq-factor}
I_{\text{p}} \equiv  \frac{p_{_{\text{Wind}}} - p_{_{\text{IRad}}}}{p_{_{\text{Wind}}} + p_{_{\text{IRad}}}},
\end{equation}
where $p_{_{\text{Wind}}}$ and $p_{_{\text{IRad}}}$ is the input of momentum from stellar winds and ionizing radiation, respectively. 
Thus, $I_{\text{p}}$ is a measure for the predominance of one feedback process and values around zero indicate equality in momentum input. We name a process ''dominant" when the relative impact is close to unity. In this analysis we assume that the sum of radiation and wind momentum input is representative for the combined momentum input. As discussed at the end of Section \ref{sec-individual}, this is a lower limit.

For each comparison, we use two simulations with identical initial conditions including either stellar winds or ionizing radiation. Fig. \ref{fig-transferefficiency_rad_wind_mom} shows the relative impact $I_{\text{p}}$ of stellar winds and ionizing radiation as a function of the source environment time. $I_{\text{p}}$ is the time-averaged value in a time period, which both simulations have in common. The vertical lines show the maximum and minimum values of $I_{\text{p}}$ obtained during the course of the simulation. Six data points are obtained from higher resolution (full markers, 0.2 pc) simulations with stars $M_{*}$ = 12 (red), 23 (blue), and 60 M$_{\odot}$ (black) embedded in the WIM and CNM (see Table \ref{tab-homogeneous1}). In addition, we include simulations with a lower uniform resolution (empty markers) of 0.4 pc in two additional, warm, neutral media, WNM$_{1}$  and WNM$_{2}$ with number densities $n_{0}$ of $\sim$ 1 and $\sim$ 10 cm$^{-3}$ and temperatures $T_{0}$ of  $\sim$ 2$\times$10$^{3}$ K and $\sim$  2$\times$10$^{2}$ K (see Table \ref{tab-lowres}).

Thin lines show the relative impact derived from the analytic estimates in Eq. \ref{eq-spitzermom} and Eq. \ref{eq-windmom} which are shifted according with the corresponding simulations. This prediction makes use of the heating and cooling balance of the ISM to relate density and temperatures.

The right ordinate shows the factor of the dominant to the subordinate process with $\frac{p_{_{\text{Wind}}}}{p_{_{\text{IRad}}}}$ above and $\frac{p_{_{\text{IRad}}}}{p_{_{\text{Wind}}}}$ below the equality of momentum, $\frac{p_{_{\text{Wind}}}}{p_{_{\text{IRad}}}} = \frac{p_{_{\text{IRad}}}}{p_{_{\text{Wind}}}}\ =\ 1$. Note that this factor diverges when $I_{\text{p}}$ approaches unity.
 
Fig. \ref{fig-transferefficiency_rad_wind_mom} reflects the results from Section \ref{sec-individual}, that stellar winds are important in the WIM and dominate radiation by a factor $\frac{p_{_{\text{Wind}}}}{p_{_{\text{IRad}}}}$ of 10$^{2}$ around a 12 M$_{\odot}$ star and up to 10$^{4}$ around a 60 M$_{\odot}$ star. In the CNM, ionizing radiation is dominant with factors $\frac{p_{_{\text{IRad}}}}{p_{_{\text{Wind}}}}$ around 50 for all considered stars. Going from the WIM to the WNM, the media change from being wind to ionizing radiation dominated. Hence, with densities larger than $n_{0}$ = 1 cm$^{-3}$ the media are radiation dominated.

The simulated and analytic values of $I_{\text{p}}$ agree in the WIM for all star masses. In environments similar to the WNM$_{1}$, a steep transition happens from the wind dominated to radiation dominated regime. The density where $I_{\text{p}}$ changes for the star with $M_{*}$ = 60 M$_{\odot}$ is smaller compared to the others because the temperature inside the H$_{\text{II}}$ region is higher for more massive stars, hence a pressure gradient establishes at lower densities. In the WNM$_{2}$ and the CNM, the analytic treatment agrees with the simulated values. In the WNM$_{1}$ and WNM$_{2}$ around the star with $M_{*}$ = 60 M$_{\odot}$, $I_{\text{p}}$ the analytical description is a factor $\sim$ 3 in $\frac{p_{_{\text{IRad}}}}{p_{_{\text{Wind}}}}$ off the simulated values. The difference arises from the assumption in the analytic model, that the temperature inside the H$_{\text{II}}$ region is not media-dependent and set to be constant. In addition, the shocks leave the computational domain early with the result of a lower momentum imposed by the ionizing radiation.

Our results disagree with the results for the low density environment discussed in the work of \citet{geen15a}. The reason is that the authors set a low temperature of 62 K in their low density environment with $n_{0}$ = 0.1 cm$^{-3}$, which disagrees with the equilibrium temperature of $\sim$ 10$^{4}$ K derived from the chemical network we employ. The authors choice of parameters enables ionizing radiation to create an over-pressured bubble and therefore overestimates the impact of radiation in their low density model.

\begin{figure}
\includegraphics[width=0.5\textwidth]{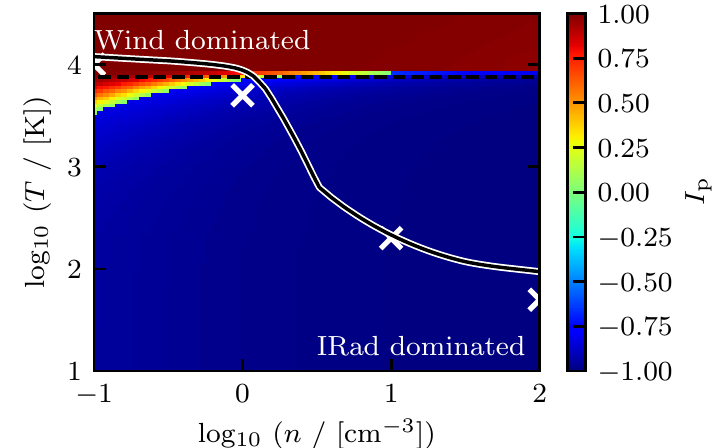} 
\caption{Time average relative impact $I_{\text{p}}$ (color) as a function of constant ambient density and temperature around a star with $M_{*}$ = 23 M$_{\odot}$. The temperature inside the corresponding H$_{\text{II}}$ region is indicated as a black, dashed line. The black, solid line corresponds to the condition of heating and cooling balance, which are the bases for the analytic predictions in Fig. \ref{fig-transferefficiency_rad_wind_mom}. The white crosses indicate the media used in this work (from left to right: WIM, WNM$_{1}$, WNM$_{2}$, CNM). }
\label{fig-mom_m23int.dat}
\end{figure}

Based on the analytic model, we can specify the media dependence. In Fig. \ref{fig-mom_m23int.dat}, we show $I_{\text{p}}$ (color) as a function of the assumed constant ambient density and temperature for a star with $M_{*}$ = 23 M$_{\odot}$. The black dashed line indicates the temperature inside the corresponding H$_{\text{II}}$ region. The black solid line shows the number density-temperature relation in equilibrium used in the analytic model described in the context of Fig. \ref{fig-transferefficiency_rad_wind_mom}. The white crosses show the ambient media assumed in this work.  Fig. \ref{fig-mom_m23int.dat} shows that above the temperature set by photoionization, the medium is wind dominated. In addition at low densities ($n < $ $\sim$ 1 cm$^{-3}$) and corresponding temperatures above 5000 K the influence of radiation decreases and $I_{\text{p}}$ approaches 1. At higher densities and lower temperatures the radiative-driven expansion dominates.

The simulations and our model assume solar metallicities. For environments with lower metallicities, the temperature inside an H$_{\text{II}}$ region is higher as metal line cooling is reduced. Hence, we expect the resulting radiative momentum to increases as well. The result would be that the wind dominated region is shifted to higher temperatures.

\section{Energy coupling and radiative cooling}
\label{sec-efficiencies}
\subsection{The coupling efficiency of stellar winds and ionizing radiation}
\label{sec-efficiency-individual}

The coupling efficiency, $\epsilon$, is a measure of how much emitted energy from a source\footnote{An additional subscript indicates the stellar feedback process, which is source of the energy, e.g. $E_{\text{src, IRad}}$}, $E_{\text{src}}$, remains in the system, $E_{\text{sys}}$, which then is able to drive radial momentum. We define $E_{\text{src}}$ as \citep{freyer03, freyer06} 
\begin{equation}
E_{\text{src}}(t)= \int_{0}^{t} L(t')\ \text{d}t',
\end{equation}
where $L$ is the source luminosity with $L = L_{\text{IRad}}$ for ionizing radiation, $L = L_{\text{Wind}}$ for stellar winds or $L = L_{\text{IRad}}+L_{\text{Wind}}$ for the combination of both processes (see Fig. \ref{fig-evolution-tracks}, bottom panel).

$E_{\text{sys}}$ is the part of the inserted energy that remains in the system in the form of kinetic and thermal energy. It is defined as
\begin{equation}
E_{\text{sys}}(t) = E(t) - E_{0}(t)
\end{equation}
where $E(t)$ is the total energy at time $t$ and $E_{0}(t)$ is the (thermal) energy of the gas in a reference box, which is evolved in isolation and slowly cooling down.

The coupling efficiency $\epsilon$ is then defined as
\begin{equation}
\epsilon \equiv \frac{E_{\text{sys}}}{E_{\text{src}}}.
\label{eq-transferefficiency}
\end{equation}

\begin{figure*}
\includegraphics[width=\textwidth]{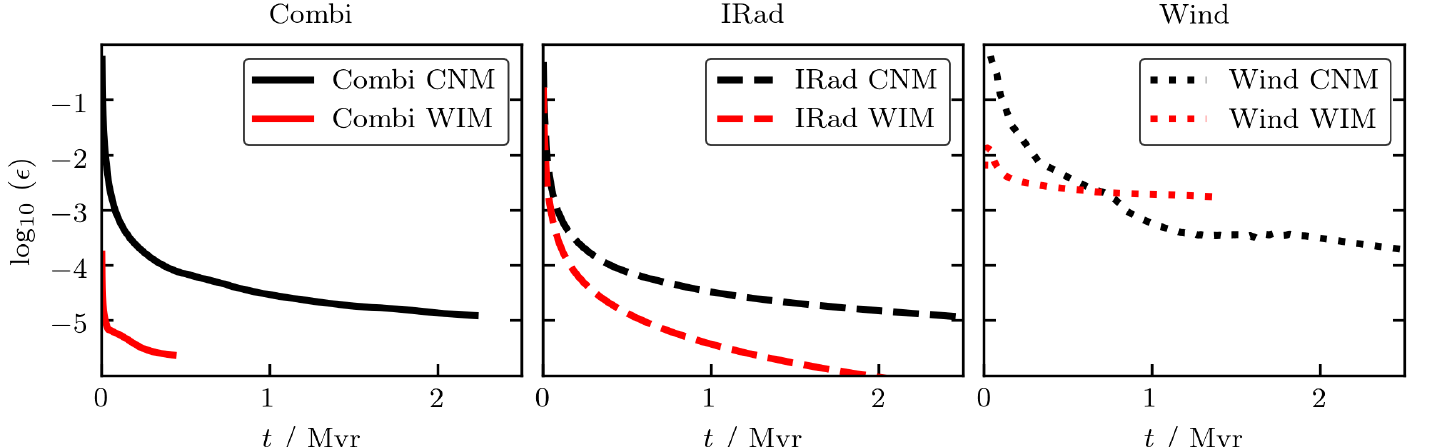} 
\caption{The evolution of the coupling efficiencies $\epsilon$ is shown for a source with 23 M$_{\odot}$. We distinguish between two media, WIM (red) and CNM (black). Ionizing radiation (middle, dashed) and stellar winds (right, dotted) are compared to the combination of both processes (left, solid). In all simulations, where the lines stop before 2.5 Myr, the feedback bubble expands out of the computational domain.}
\label{fig-transferefficiency_homogeneous_combi1}
\end{figure*}

In Fig. \ref{fig-transferefficiency_homogeneous_combi1}, we show the coupling efficiencies $\epsilon$ of ionizing radiation (middle, dashed), stellar winds (right, dotted) and the combination of both processes (left, solid) in the WIM (red) and CNM (black) for a star with a mass of 23 ${\rm M}_{\odot}$. We note that for different $M_{*}$, we find similar efficiencies and a similar time-dependent evolution. Two features are characteristic for all simulations. The initial coupling efficiency is high with $\epsilon$ = 0.9, thus almost perfect, but drops rapidly within the first 0.1 Myr. Second, stellar winds couple more efficiently to the ambient medium than ionizing radiation. 

In the CNM, the efficiency of ionizing radiation drops to $\sim$ 10$^{-5}$. For stellar winds we obtain values of $\sim$ 10$^{-4}$. However, despite the higher $\epsilon$, the wind is not important since $E_{_{\text{src, IRad}}} / E_{_{\text{src, Wind}}} \approx 10^{3}$. 
In the WIM, ionizing radiation couples to the ambient gas with $\epsilon<10^{-6}$, whereas we get $\epsilon \sim10^{-3}$ for the stellar wind. The coupling efficiency of the combination of both processes has values of $\sim 10^{-5}$, again because the radiation is more energetic than the wind but does not couple.

Overall, stellar winds couple more efficiently to the ambient medium than ionizing radiation in both, the WIM and the CNM. Ionizing radiation couples to the medium by ionizing and heating the gas, which is an inefficient process and susceptible to radiative cooling (see Section \ref{subsec-cooling}). This agrees with previous work, that show that the conversion of radiative to kinetic energy is highly inefficient (e.g. \citealt{walch12}).

\subsection{Radiative cooling}
\label{subsec-cooling}

Contrary to many previous papers, we self-consistently compute radiative heating and cooling everywhere in the computational domain. In this section, we discuss the importance of three, selected cooling processes by post-processing our simulations. These are radiative recombination of hydrogen (case B recombination, see Section \ref{sec-chemistry}), free-free emission of hydrogen, and soft X-ray emission in the energy band of 0.5 to 2 keV. The corresponding cooling rates are $\Lambda_{\text{rc}}$ \citep{cen92}, $\Lambda_{\text{ff}}$ \citep{shapiro87} and $\Lambda_{\text{X}}$. For the last rate we use tables generated with the \textsc{Astrophysical Plasma Emission Code} (APEC, \citealt{smith01}) from the collisional ionization database \textsc{AtomDB} (\citealt{foster10}, www.atomdb.org).

\begin{figure}
\includegraphics[width=0.5\textwidth]{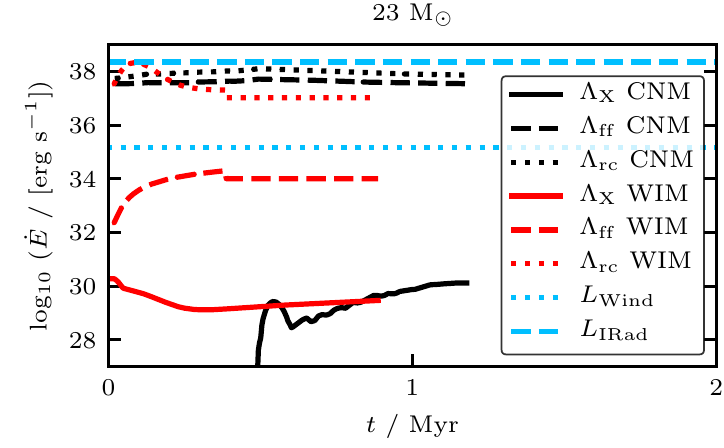} 
\caption{We compare different cooling processes in the WIM (red) and the CNM (black) when a bubble of ionizing radiation and stellar winds from a $M_{*}$ = 23 M$_{\odot}$ expands into it. We show the cooling rates associated with radiative recombination, $\Lambda_{\text{rc}}$ (dashed), free-free emission, $\Lambda_{\text{ff}}$ (dash-dotted), and X-ray emission, $\Lambda_{\text{X}}$ (dotted). For comparison we show the wind luminosity $L_{\text{Wind}}$ (blue, dotted) and the radiation luminosity $L_{\text{IRad}}$ (blue, dashed) emitted by the star. }
\label{fig-emission}
\end{figure}

In Fig. \ref{fig-emission}, we show the calculated cooling rates associated with radiative recombination (dashed), free-free emission (dash-dotted) and X-ray emission (dotted) for the two simulations where a star with $M_{*} = 23\;{\rm M}_{\odot}$ injecting a stellar wind and ionizing radiation has been placed in the WIM (red) or in the CNM (black). For comparison we show the input wind luminosity $L_{\text{Wind}}$ (blue, dotted) and the radiation luminosity $L_{\text{Irad}}$ (blue, dashed).

In the CNM, radiative recombination is the dominant cooling process with $\Lambda_{\text{rc}}\sim 9\times10^{37}$ erg s$^{-1}$ while free-free emission is a factor of 3 smaller. X-ray emission saturates at a rate of $\Lambda_{\text{X}}\sim10^{30}$ erg s$^{-1}$. About 50 percent of the total stellar luminosity, $L_{\text{src}} =L_{\text{Wind}}$+$L_{\text{IRad}}$, is lost by these three cooling processes. The residual energy is mostly lost by metal line cooling. 

In the WIM, the radiative emission with $\sim 10^{37}$ erg s$^{-1}$ is 3 orders of magnitude larger than the free-free emission, which is much smaller than the cooling rates found in the CNM. The total X-ray luminosity approaches a constant value of $\sim 10^{29}$  erg s$^{-1}$. Only $\sim 1$ percent of the total stellar luminosity is lost by the three cooling processes. 
%Metal line cooling accounts for 99 percent of the energy loss.

This indicates that the difference in coupling efficiency $\epsilon$ between ionizing radiation and stellar wind is due to cooling by radiative recombination. The interior of an H$_{\text{II}}$ region and the emissivity peak of radiative recombination have very similar temperatures (see  Eq. \ref{eq-alphab}). 

In both CNM and WIM, the energy loss by soft X-ray emission is small with $\Lambda_{\text{X}}\sim 10^{30}$ erg s$^{-1}$ or $\sim 10^{-8} L_{\text{src}}$, respectively. Therefore, the X-ray emission predicted in our model is well below the results shown by \citet{arthur07} and also below observed X-ray luminosities \citep[e.g.][]{garciasegura95a, wrigge05}. The reason is that they consider the early evolution of the wind-blown bubble (up to $\sim$ 20,000 years) where the density inside the bubble is presumably much higher and the temperature is lower, such that more soft X-ray emission is expected. Generally, X-ray emission has the peak emissivity in a temperature range of 10$^{6}$ K - 10$^{7}$ K \citep{toala16}. In the presented simulations, the X-ray emitting wind bubbles have typical temperatures of $\sim 10^8$ K (see Fig. \ref{fig-radialprofile1} and Fig. \ref{fig-app-radialprofiles}).  

\section{Application of the model to observed feedback bubbles}
\label{sec-implication}

We select two representative observed bubbles, the predominantly radiation-driven bubble RCW 120 and NGC 7635 (including the S162 complex), where wind and radiation are acting in combination. We initialize our semi-analytic model presented in the Sections \ref{sec-wind-analytic} and \ref{sec-radiation-analytic} and applied in Fig. \ref{fig-transferefficiency_rad_wind_mom} with the physical properties of these examples to estimate the relative impact of wind and radiation. The results presented in the following are rough approximations and assume spherical symmetry with homogeneous mass distribution and emission.

The H$_{\text{II}}$ region RCW 120 evolves around a $\sim$ 30 M$_{\odot}$ star of the age of 0.2 -- 0.4 Myr within an environment with densities $n_{0}$ $\sim$ 1400 -- 3000 cm$^{-3}$ \citep{zavagno07, mackey15a, figueira17}. When applying the semi-analytic model to a 30 M$_{\odot}$ star in a homogeneous medium with 1500 cm$^{-3}$ we can estimate the relative impact to be $I_{\text{p}}$ $\sim$ -0.85, indicating that radiative feedback is indeed dominant. According to our model, the star imparts $\sim$ 6.3$\times$10$^{5}$ M$_{\odot}$ km s$^{-1}$ of momentum within 2 Myr.

RCW 120 is observed in H$_{\alpha}$ and dust emission but shows no evidence of excited metal lines \citep{zavagno07}. This missing line emission indicates that the bubble is indeed radiation-driven, which agrees with our results that stellar winds are unimportant for the early evolution in the CNM. In contrast, the other, clearly wind-driven bubbles like NGC 6888 appear bright in H$_{\alpha}$ as well as metal line emission e.g [OIII]  \citep{toala12, toala13}. This evolved bubble is fully filled with wind-heated gas (compare with Fig. \ref{fig-comparison}).

An example for combined feedback is NGC 7635. The entire object emits H$_{\alpha}$ radiation while only the central part emits [OIII] in addition. From these observations and the previous discussion we can distinguish between the central, 2 pc wide, spherical, wind-driven region and the surrounding H$_{\text{II}}$ region with a diameter between 2.5 -- 3.1 pc \citep{christopoulou95, moore02}. The mean densities inside the wind-blown bubble and the H$_{\text{II}}$ region are estimated with, $n_{\text{Wind,obs}}$ = 100 cm$^{-3}$ and $n_{\text{IRad,obs}}$ = 300 cm$^{-3}$, respectively \citep{thronson82, christopoulou95, moore02, mesadelgado10}. The central source is a 0.3 Myr old, O6.5 star with a wind mass-loss rate of $\sim$~10$^{-6}$~M$_{\odot}$~yr$^{-1}$, a wind velocity of $\sim$ 2500 km s$^{-1}$, and an estimated flux of $\dot{N}_{\text{LC}}$ $\sim$ 10$^{49}$ s$^{-1}$ \citep{icke73, moore02, moore02a}, thus similar to the 23 M$_{\odot}$ star considered in this work.

The wind-blown bubble has an observed mass, $M_{\text{Wind,obs}}$, of 3 -- 14 M$_{\odot}$ and a radial expansion velocity of 4 -- 25 km s$^{-1}$ \citep{thronson82, mesadelgado10}. With this, the resulting radial momentum of the wind driven-gas, $p_{\text{Wind,obs}}$, ranges from 12 to 350~M$_{\odot}$~km~s$^{-1}$. However, for the surrounding H$_{\text{II}}$ region the corresponding properties can only be crudely estimated as measurements are insufficient. With an approximated mass, $M_{\text{IRad,obs}}$, of 26 -- 91 M$_{\odot}$ (spherical shell with $n_{\text{IRad,obs}}$) and a velocity of $\sim$ 4 -- 5 km s$^{-1}$ (expansion of 2.5 to 3.1 pc within 0.3 Myr) we obtain a momentum of $p_{\text{IRad,obs}}$ $\sim$ 110 -- 450 M$_{\odot}$ km s$^{-1}$.

We can also make use of our semi-analytic model to estimate the impact of the H$_{\text{II}}$ region: for this, we assume a 23~M$_{\odot}$ star and an environmental density of $n_{\text{IRad,obs}}$ = 300 cm$^{-3}$. The model obtains a relative impact of $I_{\text{p}}$ $\sim$ -0.9 with the total radial momentum imparted by radiation, $p_{\text{IRad,mod}}$, to be $\sim$~4700 M$_{\odot}$~km~s$^{-1}$ and the corresponding momentum from stellar winds $p_{\text{Wind,mod}}$, to be $\sim$~1300 M$_{\odot}$~km~s$^{-1}$. The modelled and observed momenta differ by up to a factor of 10. The modelled bubble radius in case of wind feedback is 2.4 pc (a factor of $\sim$ 2.4 larger than the observed radius of 1 pc) and the modelled radius in case of radiative feedback is 3.1 pc (a factor of 2 -- 2.4 larger than the observed radius with 1.25 -- 1.55 pc). These differences are closely linked to the significant uncertainties of the observed densities and velocities, the estimated mass of the central star as well as the assumption of homogeneity in the model. To obtain comparable momenta from the model assuming a 23 M$_{\odot}$ star, the ambient density has to be increased to a few 1000 cm$^{-3}$ which is comparable to the densities observed in the northern part of NGC 7635. The bubbles have expanded to the observed radii after 0.4 Myr. The relative impact is then reduced to -0.6.

Finally, we want to estimate the relative impact of feedback in NGC 7635 by comparing the observed H$_{\alpha}$ emission from the H$_{\text{II}}$ and the wind-blown region. The corresponding luminosities for the wind-driven, $L_{\text{H}\alpha\text{,Wind,obs}}$, and the radiation-driven bubble, $L_{\text{H}\alpha\text{,IRad,obs}}$, are 4.9 $\times$10$^{35}$ erg s$^{-1}$ and 1.0 -- 1.5 $\times$10$^{36}$ erg s$^{-1}$, respectively (assuming an averaged flux over a representative part of the regions, \citealt{moore02a, moore02}). To obtain the relative impact, $I_{\text{p,H}\alpha}$, we follow the idea of Eq. \ref{eq-factor} and substitute the momenta, $p_{\text{IRad}}$ and $p_{\text{Wind}}$, by $L_{\text{H}\alpha\text{,IRad,obs}}$ and $L_{\text{H}\alpha\text{,Wind,obs}}$. The resulting relative impact $I_{\text{p,H}\alpha}$ ranges from -0.4 to -0.6. This is in reasonable agreement with the estimates from the semi-analytic model, $I_{\text{p}}$ = -0.6 -- -0.9. In both cases, the relative impact in NGC 7635 is clearly dominated by radiative feedback but the wind increases in importance. We expect even better agreement when relaxing the assumptions of homogeneity (as e.g. done for supernova-driven bubbles in \citealt{haid16}).

\section{Summary}
\label{sec-summary}

In this paper, we investigate the impact and the coupling efficiencies of stellar winds and ionizing radiation from single massive stars. We perform a series of hydrodynamic simulations with the AMR code \textsc{FLASH} 4 including the novel radiative transfer scheme \textsc{TreeRay}, which is coupled to a chemical network. We use the On-The-Spot approximation with a temperature-dependent recombination coefficient and account for ionization heating within the H$_{\text{II}}$ region. The initial conditions of homogeneous ambient media match the properties of the WIM (10$^4$ K, 0.1 cm$^{-3}$, ionized), the WNM (2000 K, 1 cm$^{-3}$),  and the CNM (20 K, 100 cm$^{-3}$, predominantly neutral). Stars with $M_{*}$ = 12, 23, and 60 M$_{\odot}$ are used as sources. We expect magnetic fields, omitted in this work, to affect the shape of the feedback bubble but not change the results found here in a qualitative way.

We benchmark the coupling of the radiative transfer implementation with the chemistry module against the Monte-Carlo photoionization code {\sc MOCASSIN} and recover comparable mean temperatures within the H$_{\text{II}}$ regions for the different stars.

With ionization and stellar winds included, the radiation-driven shock is always ahead of the wind-driven shock. This implies that the wind-blown bubble is always surrounded by a H$_{\text{II}}$ region. The differential impact of stellar winds and ionizing radiation - tested separately - is highly dependent on the properties of the ambient ISM. Within the CNM, ionizing radiation dominates the momentum input ($1.6\times10^4$ to $4\times 10^5$ M$_{\odot}$ km s$^{-1}$). Stellar winds are only shock-heating a small inner bubble and contribute a negligible amount of momentum in comparison with radiation. When comparing these results to the impact of single SNe in the CNM, we find an equal or higher momentum input for stars with a mass of 23 M$_{\odot}$ and above. In the WIM, the momentum input of stellar winds is similar to the CNM ($2\times 10^2$ to $5\times 10^3$ M$_{\odot}$ km s$^{-1}$), while ionizing radiation falls short ($\sim 10^2 $ M$_{\odot}$ km s$^{-1}$). With both processes at work, ionizing radiation supports the wind-driven expansion by preventing the rarefied environment from cooling and recombining. We also show that the warm neutral medium is a transition regime from ionization dominated momentum injection in the CNM to wind dominated injection in the WIM. We introduce an analytic model to predict in which homogeneous media ionizing radiation or stellar wind is dominating. The values from this description are similar to the results from the numerical simulations. 

Energetically, stellar winds couple more efficiently to the ISM ($\sim$ 0.1 percent of wind luminosity) than ionizing radiation ($<$ 0.001 percent of ionizing luminosity). The low coupling efficiency of ionizing radiation results from the high cooling rate associated with radiative recombination and free-free emission. 

For estimating the global impact of massive stars on different surrounding media, the strongly mass-dependent ratios of wind luminosity to ionizing luminosity (see Fig. \ref{fig-evolution-tracks}) have to be considered. It is likely, that a massive star interacts with vastly different environments during its lifetime due to the short dispersal time scales of young star clusters and the significant fraction of runaway massive stars. In summary, this study shows that the relative impact of stellar winds and ionizing radiation depends on the stellar mass considered but even more strongly on the properties of the ambient medium.

\section{Acknowledgements}

SH, SW, DS and FD acknowledge the support by the Bonn-Cologne Graduate School for physics and astronomy which is funded through the German Excellence Initiative.  SH, SW, and DS also acknowledge funding by the Deutsche Forschungsgemeinschaft (DFG) via the Sonderforschungsbereich SFB 956 ''Conditions and Impact of Star Formation" (subproject C5). SH, SW, DS, FD and TN acknowledge the support by the DFG Priority Program 1573 ''The physics of the interstellar medium". SH and SW acknowledge funding by the European Research Council through ERC Starting Grant No. 679852 ''RADFEEDBACK". TN acknowledges support from the DFG cluster of excellence ''Origin and Structure of the Universe". R.W. acknowledges support by the Albert Einstein Centre for Gravitation and Astrophysics via the Czech Science Foundation grant 14-37086G and by the institutional project RVO:67985815 of the Academy of Sciences of the Czech Republic. The software used in this work was developed in part by the DOE NNSA ASC- and DOE Office Science ASCR-supported \textsc{FLASH} Center for Computational Science at University of Chicago. The authors gratefully acknowledge the Gauss Centre for Supercomputing e.V. (www.gauss-centre.eu) for funding this project by providing computing time on the GCS Supercomputer SuperMUC at Leibniz Supercomputing Centre (www.lrz.de). We thank the \textsc{YT-PROJECT} community \citep{turk11} for the \textsc{YT} analysis package, which we used to analyse and plot most of the data. We thank the anonymous referees for the constructive input.

\bibliographystyle{aa}
\bibliography{GeneralBib}

\appendix

\section{Interstellar radiation field}
\label{sec-G0}
\begin{figure}
\includegraphics[width=0.5\textwidth]{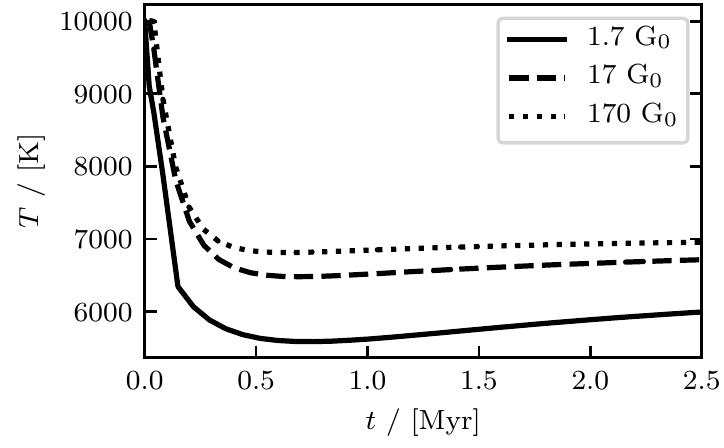} 
\caption{Temperature evolution of the WIM without a massive star under the conditions of a homogeneous ISRF with G$_{0}$ = 1.7 (solid), 17 (dashed) and 170 (dotted).}
\label{fig-G0}
\end{figure}

As discussed in Section \ref{sec-chemistry}, we include a background interstellar radiation field (ISRF) of homogeneous strength G$_{0}$ = 1.7 \citep{habing68, draine78}. To assume the ISRF to be constant near a massive star is an approximation. In Fig. \ref{fig-G0}, we show the influence of the ISRF on the WIM, hence the temperature evolution. We expect the FUV radiation to penetrate deeper into the rarefied medium and choose three homogeneous strength of G$_{0}$ = 1.7 (solid), 17 (dashed) and 170 (dotted). The initial drop of temperature is slowed down as the ISRF is increased. However, the maximum possible temperature difference is 20 percent with an average difference of under 10 percent between the G$_{0}$ = 1.7 and 170. We conclude, that a homogeneous ISRF is valid for the presented simulations. 

\section{Radial profiles}
\label{sec-app-radial}
\begin{figure*}
\includegraphics[width=\textwidth]{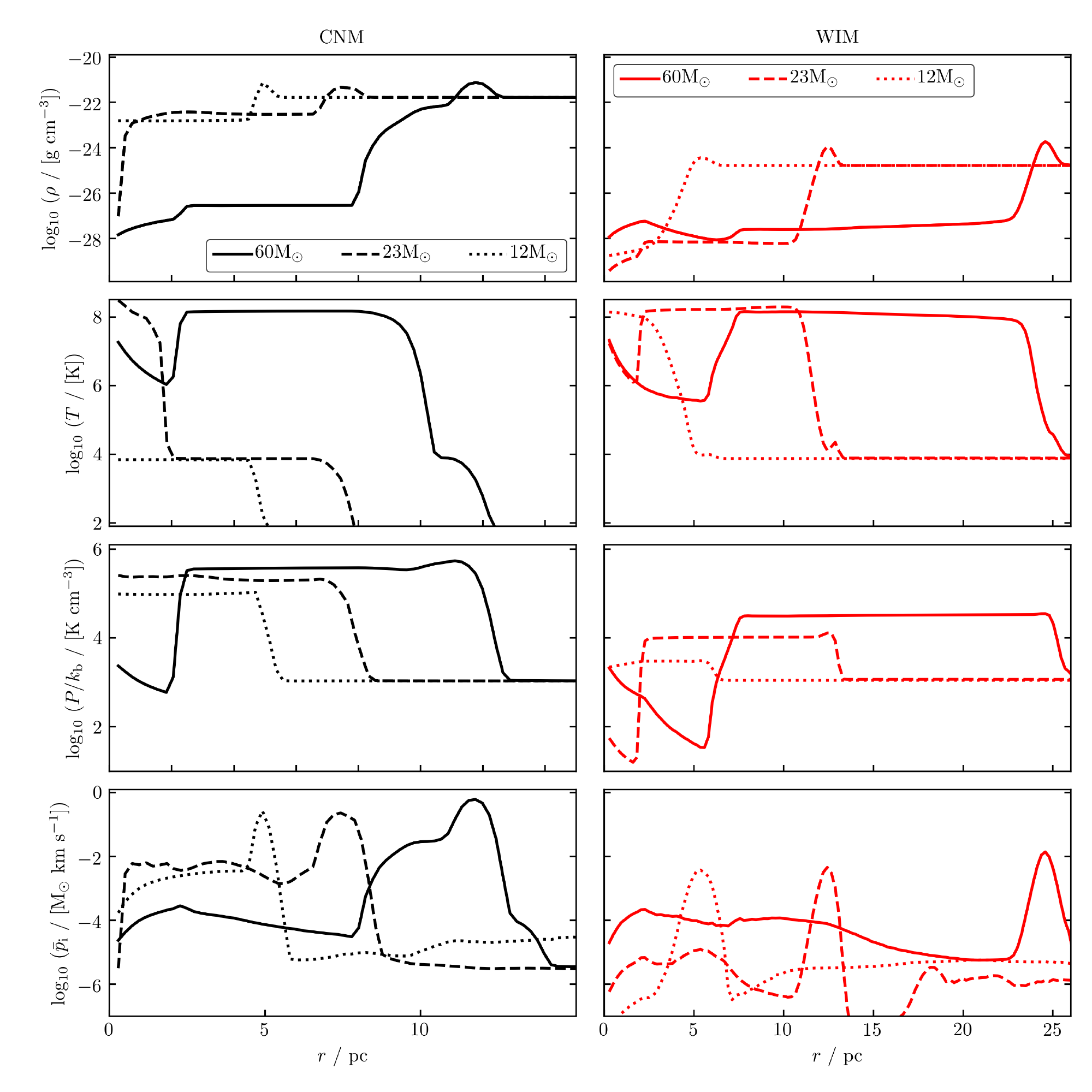} 
\caption{The radial profiles of the combination of stellar winds and ionizing radiation which were obtained from the simulation for Fig. \ref{fig-comparison}. The left (red) and the right (black) column shows the radially averaged values in the WIM and the CNM. Each panel includes the structure from stars with $M_{*}$ = 12 (dotted), 23 (dashed) and 60 (solid) M$_{\odot}$. For top to bottom we show the density $\rho$, the temperature $T$, the pressure over the Boltzmann constant $P/k_{\text{B}}$ and the mean momentum of a cell i $\bar{p}_{\text{i}}$. The x-axis match the zoom-in length scale of Fig. \ref{fig-comparison} with 25 and 15 pc.}
\label{fig-app-radialprofiles}
\end{figure*}

In Fig. \ref{fig-app-radialprofiles}, we show the radial profiles of density $\rho$ (top), the temperature $T$ (second), the pressure over the Boltzmann constant $P/k_{\text{B}}$ (third) and the mean momentum of a cell i $\bar{p}_{\text{i}}$ (bottom) for simulations with stellar winds and ionizing radiation in the CNM (left) and the WIM (right) for stars with $M_{*}$ = 12 (dotted), 23 (dashed) and 60 (solid) M$_{\odot}$.

In the CNM, the shock position increases with increasing mass of the star. The behaviour of density and temperature are discussed in Section \ref{sec-combi-radial}. The third panel shows, that the expansion is pressure driven with a pressure contrast of almost 2 orders of magnitude. The resulting shock carries most of the radial momentum. The interior expands too but with a significant lower momentum. 

In the WIM, the shock density is a factor of $\sim$ 4 higher than the ambient density. Within the almost homogeneous interior the density drops to $\sim$ 10$^{-28}$ g cm$^{-3}$ with a temperature of $\sim$ 10$^{8}$ K. The result are pressures between $\sim$ 3 $\times$ 10$^{3}$ K cm$^{-3}$ and $\sim$ 2 $\times$ 10$^{4}$ K cm$^{-3}$ for the star with $M_{*}$ = 12 and 60 M$_{\odot}$, respectively. The wind-driven shock contains a significant fraction of the radial momentum. The momentum in the shock is 3 orders of magnitude higher compared to the interior.

\section{Impact of radiation pressure}
\label{sec-radpres}
\begin{figure}
\includegraphics[width=0.5\textwidth]{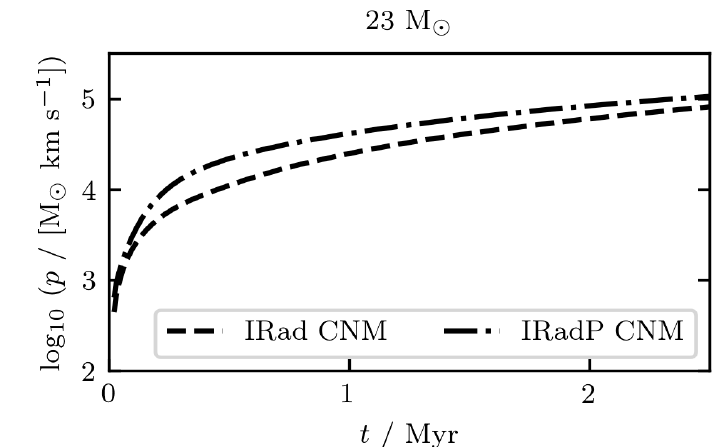} 
\caption{Momentum evolution of ionization feedback from a 23 M$_{\odot}$ source in the CNM. We compare the radiation feedback without (dashed) and with (dash-dotted) radiation pressure.}
\label{fig-comparison-rp}
\end{figure}

The impact of radiation pressure is highly debated. This process is considered to be unimportant in the CNM \citep{rahner17} and with a source luminosity $L_{\text{IRad}}$ from a single star with $M_{*} \approx 23\;{\rm M}_{\odot}$ \citep{krumholz09c,fall10, murray11,sales14}.

In Fig. \ref{fig-comparison-rp}, we compare the momentum evolution from ionizing radiation feedback with (dash-dotted) and without radiation pressure (dashed) on gas in the CNM. At $t$ = 2.5 Myr, the momentum input from the radiation only simulation is 8.4$\times$10$^{4}$ M$_{\odot}$ km s$^{-1}$ and increases by about 20 percent with additional radiation pressure. We conclude that radiation pressure is subdominant. In the WIM, the impact of radiation pressure is expected to be even smaller, due to the inefficient coupling of the radiation.

\end{document}